\documentclass[10pt,preprint]{aastex}

\newcommand{\kms}{\,{\rm km\,s^{-1}}}
\newcommand{\msun}{\,{\rm M_\odot}}

\newcommand{\etal}{{et al.\ }}

\def\spose#1{\hbox to 0pt{#1\hss}}
\def\lta{\mathrel{\spose{\lower 3pt\hbox{$\mathchar"218$}} \raise 2.0pt\hbox{$\mathchar"13C$}}}
\def\gta{\mathrel{\spose{\lower 3pt\hbox{$\mathchar"218$}} \raise 2.0pt\hbox{$\mathchar"13E$}}}

\begin{document}
\title{Simulations of Recoiling Massive Black Holes in the Via Lactea Halo}
\author{J. Guedes\altaffilmark{1}, P. Madau\altaffilmark{1}, M. Kuhlen\altaffilmark{2},
J. Diemand\altaffilmark{1,3}, and M. Zemp\altaffilmark{4}}

\altaffiltext{1}{Department of Astronomy \& Astrophysics, University of California,
Santa Cruz, CA 95064, USA}
\altaffiltext{2}{Institute for Advanced Study, Einstein Drive, Princeton, NJ 08540, USA}
\altaffiltext{3}{Hubble Fellow.}
\altaffiltext{4}{Astronomy Department, University of Michigan, Ann Arbor, MI 48109, USA}

\begin{abstract}

The coalescence of a massive black hole (MBH) binary leads to the
gravitational-wave recoil of the system and its ejection from the galaxy core.
We have carried out $N$-body simulations of the motion of a $M_{\rm BH}=3.7 
\times 10^6\,\msun$ MBH remnant in the ``Via Lactea I'' simulation, a Milky Way sized 
dark matter halo. The black hole receives a recoil velocity of $V_{\rm kick}$ = 80, 120, 
200, 300, and 400 $\kms$ at redshift 1.5, and its orbit is followed for over 1 Gyr within 
a ``live" host halo, subject only to gravity and dynamical friction against the dark matter 
background. We show that, owing to asphericities in the dark matter potential, the orbit 
of the MBH is highly non-radial, resulting in a significantly increased decay timescale 
compared to a spherical halo. The simulations are used to construct a semi-analytic 
model of the motion of the MBH in a time-varying triaxial Navarro--Frenk--White dark 
matter halo plus a spherical stellar bulge, where the dynamical friction force is calculated 
directly from the velocity dispersion tensor. Such a model should offer a realistic picture 
of the dynamics of kicked MBHs in situations where gas drag, friction by disk stars, and 
the flattening of the central cusp by the returning black hole are all negligible effects. We find 
that MBHs ejected with initial recoil velocities $V_{\rm kick}\gta 500\,\kms$ do not return to the 
host center within a Hubble time. In a Milky Way-sized galaxy, a recoiling hole 
carrying a gaseous disk of initial mass $\sim M_{\rm BH}$ may shine as a quasar  for a 
substantial fraction of its ``wandering" phase. The long decay timescales of kicked 
MBHs predicted by this study may thus be favorable to the detection of off-nuclear 
quasar activity. 

\end{abstract}


\section{Introduction}

Intermediate-mass black holes may have formed at redshift $z\gta 15$ at the bottom 
of shallow dark-matter potential wells \citep{madau01}. These seed holes may have 
grown through gas accretion and binary coalescences to become the supermassive 
variety that is ubiquitously found today at the center of nearby galaxies 
\citep{kormendy95,richtsone98, tremaine02}. In the context of cold dark matter (CDM) 
cosmologies, where large halos are assembled via the hierarchical assembly and 
accretion of smaller progenitors, close MBH binaries inevitably form in large numbers 
during cosmic history \citep{begelman80,volonteri03}. The  presence of a MBH binary 
with separation $<1\,$kpc has been revealed by {\it Chandra} observations of the nucleus of NGC 6240 
\citep{komossa03,max07}.  

The Very Long Baseline Array (VLBA) discovery in the radio galaxy 0402+379 of a MBH binary system with a 
projected separation of just 7 pc and a combined mass of $\sim 1.5\times
10^8\,\msun$ was reported  by \cite{rodriguez06}. A 
MBH binary may shrink owing to stellar and/or gas dynamical processes 
\citep[e.g.,][]{mayer07} and finally coalesce when gravitational wave radiation 
dominates orbital energy losses. 

Recent developments in numerical relativity \citep{pretorius05,campanelli06,
baker06a} have allowed several groups to simulate the coalescence phases
of black hole binaries \citep{baker06a,herrmann07,gonzalez07a}. 
Gravitational wave emission is typically anisotropic because of asymmetries 
associated with the masses and spins of the black holes, and causes the 
center of mass of the system to recoil in order 
to balance the linear momentum carried away by gravitational radiation 
\citep{bekenstein73,fitchett84,favata04}. 
The recoil velocity $\vec{V}_{\rm kick}$ depends on the binary mass ratio $q_b=M_1/M_2<1$ on the
dimensionless spin vectors of the pair $\vec{a}_1$ and $\vec{a}_2$ ($0<a_i<1$), and on
the orbital parameters. All current numerical data on kicks can be fitted by \citep{baker08}
\begin{eqnarray}
\vec{V}_{\rm kick} &=& v_m \, \vec{e}_x + v_{\perp} (\cos\xi \, \vec{e}_x + \sin\xi \, \vec{e}_y)
+ v_{\parallel} \, \vec{e}_z, \label{eq:v_total}\\
      v_m     &=& A \mu^2 \sqrt{1 - 4 \mu}\, (1 + B \mu), \label{eq:v_mass}\\
v_{\perp}     &=& H \mu^2(1+q_b)^{-1}\left( a_2^{\parallel} - q_b a_1^{\parallel}
\right), \label{eq:v_perp}\\
v_{\parallel} &=& K\mu^3(1+q_b)^{-1} [q_b a_1^{\perp}\cos(\phi_1 - \Phi_1) - a_2^{\perp}\cos(\phi_2-\Phi_2)],  \label{eq:v_parallel}
\end{eqnarray}
where $\mu=q_b/(1+q_b)^2$ is the symmetric mass ratio, $\theta_i$ is the angle 
between the dimensionless spin vector $\vec{a}_i=\vec{S}_i/M_i^2$ of the $i$th black hole and 
orbital angular momentum vector, $\phi_i$ is a projection angle between the spin 
vectors and a reference angle that lies in the orbital plane, and $\Phi_1(q_b) = \Phi_2(1/q_b)$ 
are constant for a given value of $q_b$. Here, $A = 1.35 \times 10^4 \kms$, $B = -1.48$, $H = 7540 \pm 160 \kms$, $\xi =
215^\circ \pm 5^\circ$, and $K = 2.4 \pm 0.4 \times 10^5 \kms$.
Assuming random spin orientations, $q_b>1/4$, and $a_1=a_2=0.9$, recoiling black 
holes can get a kick velocity $>500\,\kms$ approximately 60\% of the time \citep[see Table 3 of ][]{baker08}. 
For $q_b>0.1$, the percentage of kicks with $>500\,\kms$ decreases to $\sim 20\%$. 
Spins that are aligned with the orbital angular momentum vector
(as expected under the action of external torques provided by a circumbinary accretion
flow, see \citealt{bogdanovic07}) yield recoil velocities below $200\,\kms$,
while the configuration producing the maximum recoil kick corresponds to equal-mass maximally 
rotating holes with anti-aligned spins oriented parallel to the
orbital plane, $V_{\rm kick}=K\mu^3=3750\,\kms$ \citep{campanelli07a}.  

If not ejected from the host altogether,
the recoiling MBH will travel some maximum distance and then return to the center
subject to dynamical friction \citep{madau04}. Galaxy mergers are also a
leading mechanism for supplying gas to their nuclear MBHs, and a recoiling hole
can retain the inner parts of its accretion disk, providing fuel for a continuing luminous
phase along its trajectory. Two possible observational manifestations of
gravitational-radiation ejection have then been suggested: (1) spatially offset active galactic nuclei (AGN) activity
\citep{madau04,blecha08,volonteri08}; and (2) broad emission lines that are substantially 
shifted in velocity relative to the narrow-line gas left behind \citep{bonning07}.
The effect of gravitational wave recoil in the mass buildup of MBHs is more prominent at high redshifts  
\citep[e.g.,][]{volonteri06,tanaka09} and therefore the detection of offset nuclei is difficult. 
Observational evidence of recoiling MBHs is scarce and highly controversial. A recoiling SMBH 
candidate at z = 0.71 was reported by \cite{komossa08} in quasar SDSS J092712.65+294344.0. The broad-line region of the quasi-stellar object (QSO), powered by a $6\times 10^8 M_{\odot}$ black hole, appeared to have a velocity offset of $2650\,\kms $ with respect to the narrow-line region associated with the galaxy. However, several authors have challenged this hypothesis, proposing that  the object is a MBH binary \citep{dotti08,bogdanovic09} or an interacting galaxy pair \citep{shields09,heckman09}.

The observability of recoiling MBHs depends sensitively on their dynamics in galaxy halos. 
The radial orbit of a recoiling hole in a spherically symmetric potential was first 
studied analytically by \cite{madau04} and numerically by \cite{boylan-kolchin2004}. 
These early studies showed that large kicks ($\sim 400\,\kms$) can 
displace MBHs tens of kiloparsecs away from the center of a Milky Way-sized stellar bulge
and that, after the kick, the MBH undergoes several oscillations 
before decaying back to the bottom of the potential. Most of the orbital 
energy is lost during the MBH passages through the center, where dynamical
friction is most efficient: the cuspy central stellar density profile is flattened by the heating 
effect of dynamical friction, and the MBH decay timescale correspondingly lengthened.
\cite{gualandris08} have recently substantiated these results by performing direct summation 
$N$-body simulations of MBH recoil in spherical galaxies with binary-depleted cores. They 
found that initially the MBH loses its energy due to dynamical friction as predicted by Chandrasekhar's 
theory \citep{chandrasekhar43}. When the amplitude of the motion has fallen to roughly the core radius, 
the MBH and core experience damped oscillations about their common center of mass, which decay 
until the hole reaches thermal equilibrium with the surrounding stars. \cite{vicari07} evaluated the effect 
of non-spherical galaxy geometries on kicked MBHs using triaxial models, and found significantly longer 
decay timescales than in equivalent spherical systems, as in a non-spherical potential the hole does 
not return directly through the dense center where the dynamical friction force is highest. 
Blecha \& Loeb (2008) studied the trajectories of kicked holes in a two-component galaxy model that 
includes a spherical stellar bulge and a gaseous disk, and found that kicks with initial velocity 
$V_{\rm kick} \lta 200\,\kms$ in the plane of the disk are quickly damped out in $t \lta 10^{6.5}$ yr. 

In this paper, we revisit the problem using a different approach. We carry out full 
$N$-body simulations of a recoiling MBH that is subject only to gravity and dynamical 
friction against the dark matter background, in a high-resolution, 
non-axisymmetric, ``live" potential. The host is the main halo of the {\it Via Lactea I} 
(hereafter VL-I) cosmological simulation (Diemand et al. 2007a, 2007b). We follow the MBH 
orbital behavior starting at redshift $z=1.5$ (when the kick is assumed to occur) for more 
than 1 Gyr, as the host grows in mass and changes its shape from prolate to triaxial. We show that,
owing to departures from axisymmetry in the dark matter potential, the orbit of the hole is highly 
non-radial, resulting in a significantly increased decay timescale compared to a spherical halo.
The simulations are used to construct a more realistic semi-analytic model of the motion
of the MBH in a time-varying triaxial Navarro--Frenk--White (NFW) halo plus a fixed isothermal
stellar bulge, where the dynamical friction force is calculated directly from the velocity 
dispersion tensor.  Such a model should offer a more realistic
picture of the dynamics of kicked MBHs in situations where gas drag, friction by disk
stars, and the heating effect of the returning hole on the central cusp are all negligible.

\section{Simulations setup and properties of the host}

The VL-I simulation was performed with PKDGRAV (Stadel 2001) a cosmological tree code that 
includes gravitational multipoles up to hexa-decapole order to reach high accuracy in the force 
calculation. It employed multiple mass particle grid initial conditions generated with the GRAFIC2 
package \citep{bertschinger01} in a {\it WMAP} 3-year cosmology \citep{spergel07}. 
A bug in the original GRAFIC2 code caused the power spectrum used for the 
VL-I refinements to be that of the baryonic component, equivalent to 
an effective spectral index of $n=0.90$ instead of the intended 0.95. 
In this cosmology subhalo concentrations and peak circular velocities are slightly 
lower than in {\it WMAP} 3-year, while $\sigma_8$ and the main halo properties 
remain the same.\footnote{Note that this problem does not affect the more recent ``Via Lactea 
II'' and ``GHALO" simulations \citep{diemand08,stadel08}.}~ The high-resolution region was 
sampled with 234 million particles of mass $m_p = 2.1\times 10^4\msun$ and evolved
with a force resolution of $\epsilon=90$ pc. It was embedded within a periodic
box of comoving size $L=90$ Mpc, which was sampled at lower resolution to
account for the large-scale tidal forces.  The host halo mass at $z=0$ is $M_{200}=1.8\times 
10^{12}\,\msun$ within a radius of $R_{200}=389$ kpc (defined as 
the radius within which the enclosed average density is 200 times the mean matter value). 
In this work we rerun VL-I using PKDGRAV from redshift $z_i=1.54$ to $z_f=1.15$, and follow the orbits
of all dark matter particles as well as a new MBH particle placed at the center of the host. 
As in the original VL-I simulation, we employ a gravitational softening of 90 pc for the dark matter 
particles and the MBH, as well as adaptive time steps as short as $\tau = 68,500$ yr, 
sufficient to ensure convergence in the density profile down to a radius of $r_{\rm conv} \sim1.0$ kpc and 
to accurately sample the orbit of the MBH. The time-stepping criterion is given by $\Delta t <0.2 \sqrt{\epsilon/a_l}$, 
where $a_l$ is the local acceleration.  The resolution of VL-I allows us to adopt the mass of SgrA*,  
$M_{\rm BH} = 3.7 \times 10^6\,\msun$ \citep{ghez05}, for the central MBH particle: this implies a MBH-to-particle 
mass ratio of 175, enough to accurately reproduce the effect of dynamical friction.  

Large kicks can displace MBHs sufficiently far away that their decay times become 
a significant fraction of the age of the universe. It is interesting to look at the 
evolution of the host halo in terms of its time-varying spherically averaged density 
profile and shape parameters. The fitting formula proposed by \citet{navarro97}
provides a reasonable approximation to the density profile, 
\begin{equation}
  \rho(x) = \frac{\rho_s}{x(1 + x)^2},  
\end{equation}
where $x=r/R_s$ and $R_s$ is the scale radius. The mass profile is given by $M(<x)=M_{200}
f(x)/f(c)$, where $f(x)\equiv \ln(1+x)-x/(1+x)$ and $c \equiv R_{200}/R_s$ is the concentration 
parameter. The escape speed from the halo center is
\begin{equation}
v_{\rm esc}^2(0) = 2 \int_0^{\infty}\frac{G M(<r)}{r^2}\,dr = \frac{2 V_{200}^2 c}{f(c)},
\end{equation}
where $V_{200}^2 \equiv GM_{200}/R_{200}$. The quantities $\rho_s, R_s, R_{200}, M_{200}$, 
$V_{\rm max}$ (the maximum circular velocity of the host) and $v_{\rm esc}(0)$ are given in 
Table 1  at different scale factors, starting with the time when the kick is imparted. 

\begin{figure}[thb]
\centering
\includegraphics[width=0.6\textwidth]{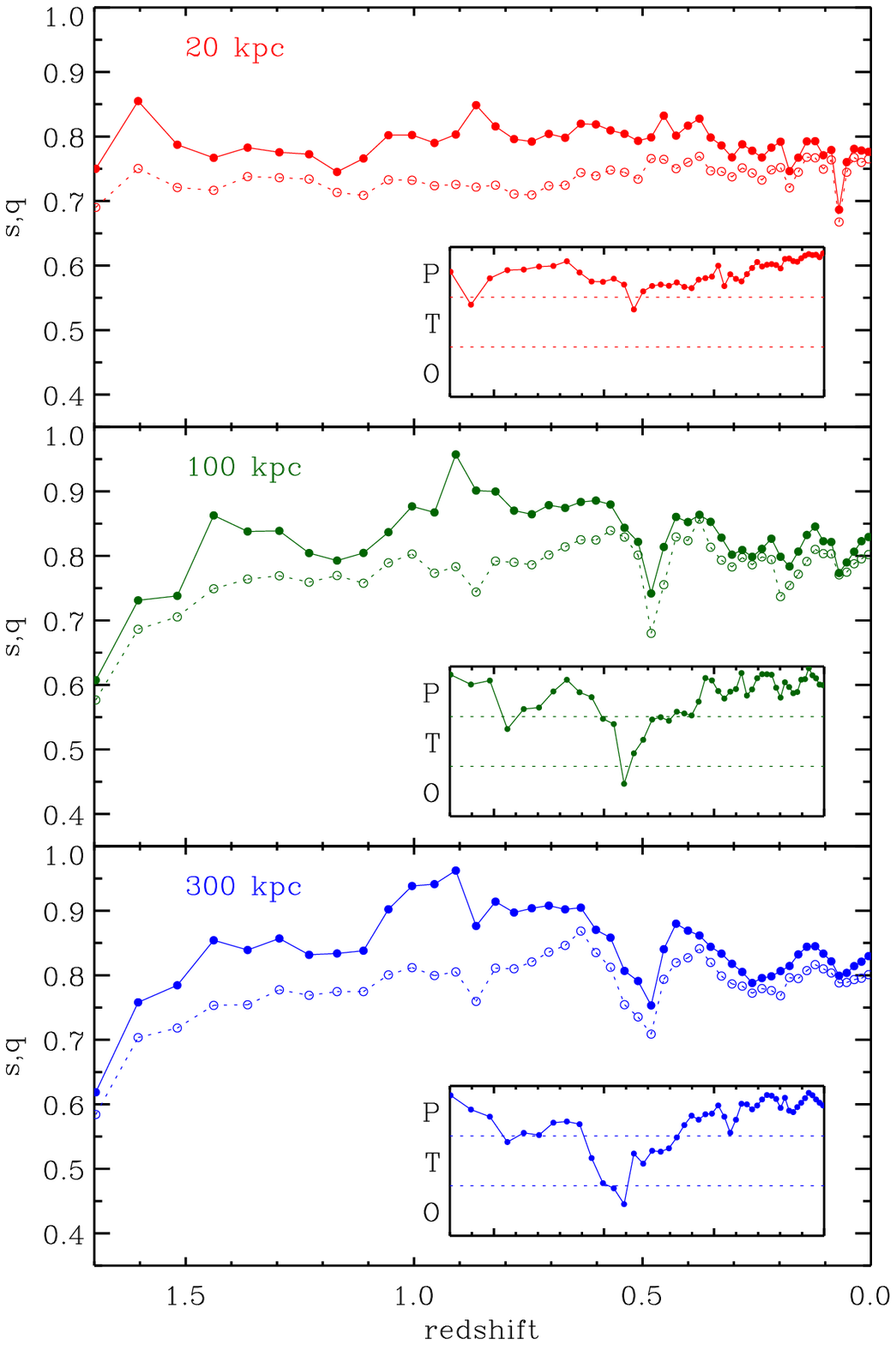}
\caption{\footnotesize Host halo (potential) shape parameters as a function 
of redshift at different ellipsoidal radii. Intermediate-to-major axis ratio $q$
(\textit{solid points}), minor-to-major axis ratio $s$ (\textit{empty circles}), and
triaxiality parameter $T$ (\textit{insets}).} Insets have the same x-axis range as the main plots. 
\label{shapes}
\end{figure}

CDM halos are known to show significant departures from sphericity (for a recent summary, see
Allgood \etal 2006). As detailed in \citet{kuhlen07}, we approximate the shape of the VL-I host potential 
by diagonalizing the unweighted kinetic energy tensor
\begin{equation}
K_{ij} = \frac{1}{2} \sum_n v_{i,n} v_{j,n},
\end{equation}
where $K_{ij}$ is related to the potential energy tensor
$W_{ij} = \sum x_i d\Phi/dx_j$ through the tensor virial theorem
\begin{equation}
\frac{1}{2} \frac{d^2I_{ij}}{dt^2} = 2 K_{ij} + W_{ij}.
\end{equation}
Here,  
\begin{equation}
I_{ij} = \sum_n \frac{x_{i,n} x_{j,n} }{r^2_n}.
\end{equation}
and $r_n = \sqrt{(x_n^2+(y_n/q)^2+(z_n/s)^2)}$. We assume $d^2I_{ij}/dt^2=0$ so that the eigenvectors of $K_{ij}$ reflect the
principal axes of the potential ellipsoid. The latter is significantly rounder than 
the mass distribution, and neither its shape nor orientation varies much with the radius 
\citep{kuhlen07}. The degree of triaxiality of the halo potential, $T$, is given by 
\citep{franx91}
\begin{equation}
T = \frac{1-q^2}{1-s^2},
\end{equation}
where $q=b/a$ and $s=c/a$ are the time-dependent intermediate-to-major and minor-to-major 
axis ratios, respectively ($a \geq b \geq c$). A halo is said to be oblate for $T<1/3$, 
triaxial for $1/3<T<2/3$, and prolate for $T>2/3$. Figure~\ref{shapes} shows the 
evolution of the potential shape parameters with redshift at different radii.
In the inner regions the axis ratios remain approximately constant after around $z=0.8$, 
but before $z=1$ there are significant changes in the outer regions, as the
halo becomes more spherical. The triaxiality parameter remains mostly in the prolate 
regime ($>2/3$) in the inner regions, while in the outer halo evolves from prolate at $z\gta 1$ 
to triaxial or slightly oblate at $0.7\lta z \lta 1$, to back to prolate at later times.
Note that the VL-I host accretes some fairly massive subhalos between $z=1$
and $z=0.5$. Dynamical friction causes these subhalos to spiral in to
the center over a few orbits, and they lose most of their mass in this 
process. The associated redistribution of material probably contributes 
to the observed shape adjustments.

\section{Dynamics of recoiling holes}

\subsection{Orbits in numerical simulations}\label{simulations}

We placed the MBH particle at the position of the densest point of the main VL-I  halo at 
an initial redshift $z_i = 1.54$, 300 Myr after the last major merger. At this epoch the host has 
$M_{200} = 1.02 \times 10^{12}\,\msun$ and $R_{200} = 187$ kpc. The kick was oriented 
at an angle of 20$^\circ$ to the minor axis of the host halo at $z_i$. The MBH orbit was 
tracked at every time step in our simulations, and its position and velocity were measured 
with respect to the central position and center of mass velocity, respectively. The five 
halo$+$MBH runs ---corresponding to kick velocities $V_\mathrm{kick}$ = 80, 120, 200, 
300, and 400 $\kms$ and labeled VL080 to VL400---  were evolved for 1.15 Gyr (i.e., until 
a final redshift $z_f=1.15$).  All kick velocities are below the escape speed at $z_i$, 
$v_{\rm esc}(r=0,z_i)=488\,\kms$. Each run consumed 13,000 CPU hours on the Pleiades 
Supercomputer Cluster at UCSC, and followed the MBH for 10,000 time steps.

\begin{figure}[thb]
\centering
\includegraphics[width=0.7\textwidth]{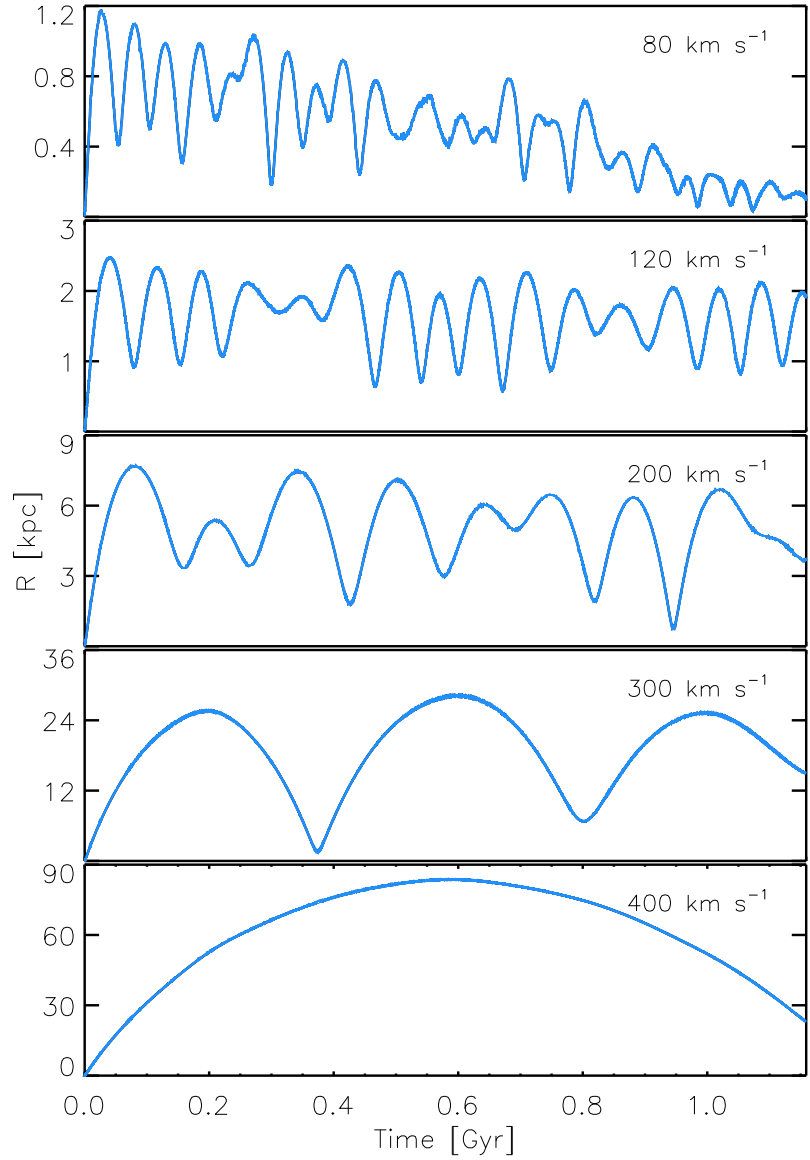}
\caption{\footnotesize Response of a  $M_{\rm BH}=3.7 \times 10^6\,\msun$ MBH to a kick at $z_i=1.54$ 
in the aspherical potential of the ``live'' VL-I Milky Way-sized halo. The radial distance $R$ of the 
hole from the center is plotted vs. time for $V_\mathrm{kick}$ = 80, 120, 200, 300, and 400 $\kms$.
Each orbit was sampled with 10,000 points. }
\label{all_runs}
\vskip 0.3cm
\end{figure}

\begin{figure}[thb]
\centering
\includegraphics[width=0.45\textwidth]{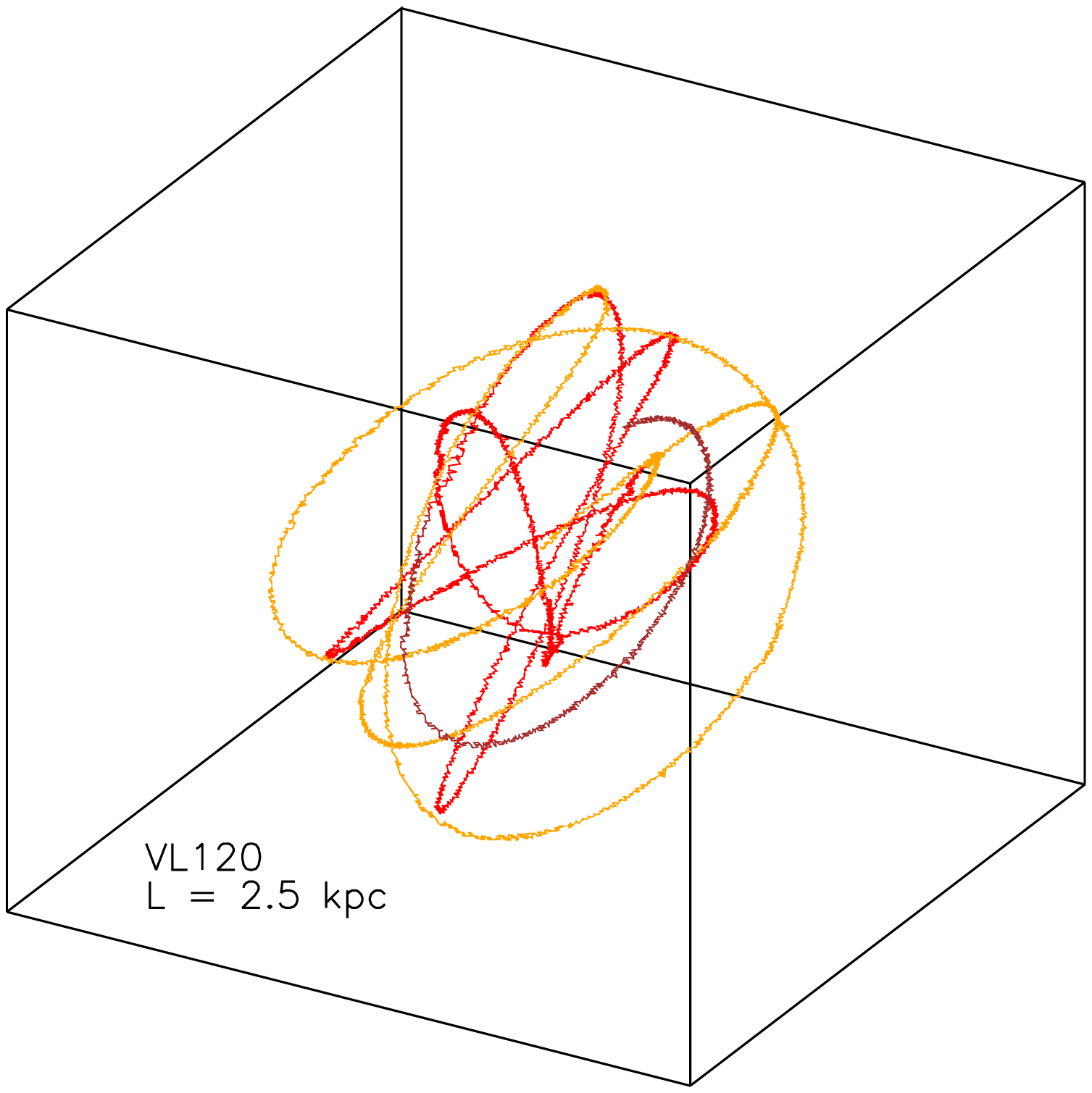}
\includegraphics[width=0.45\textwidth]{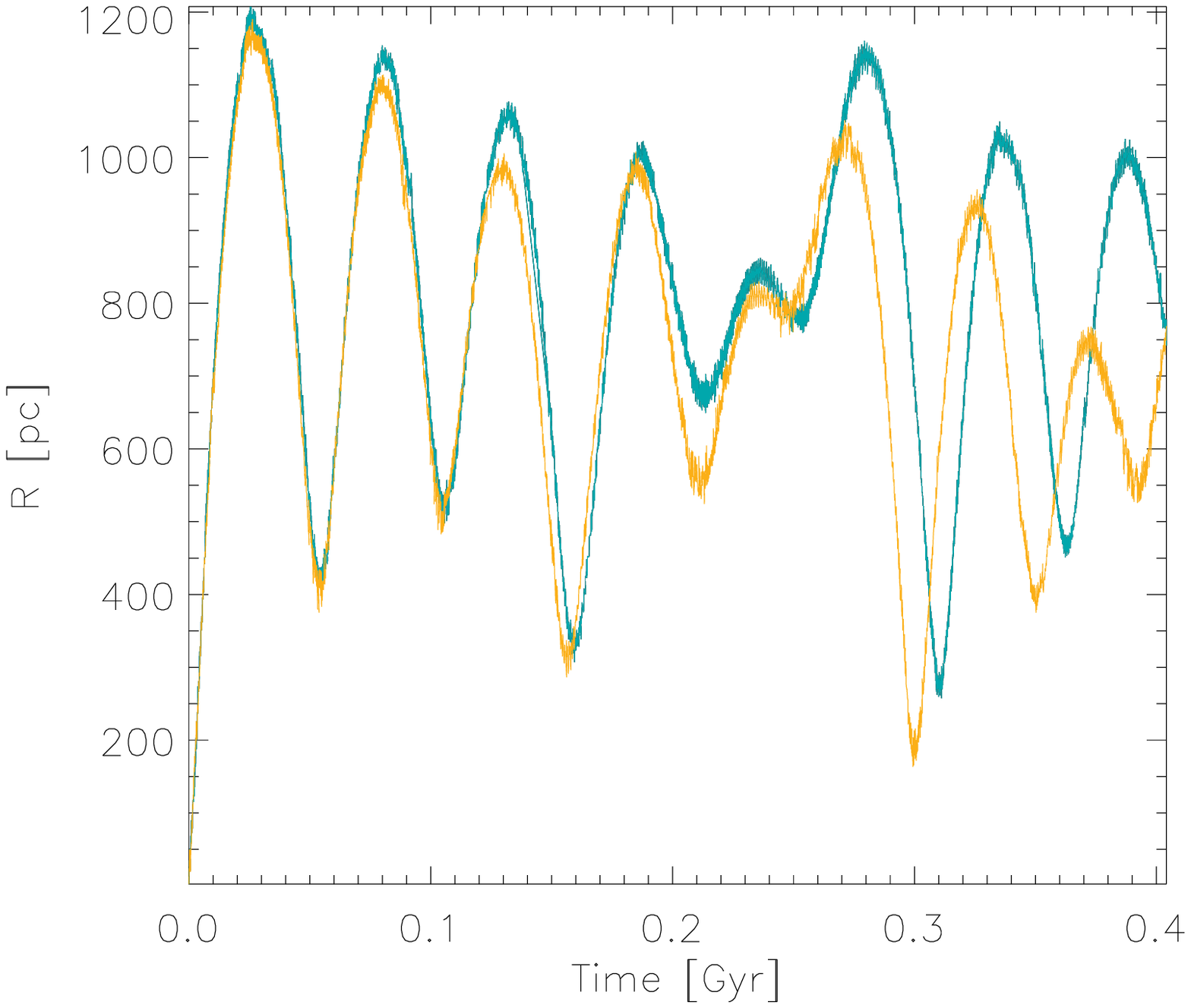}
\includegraphics[width=0.45\textwidth]{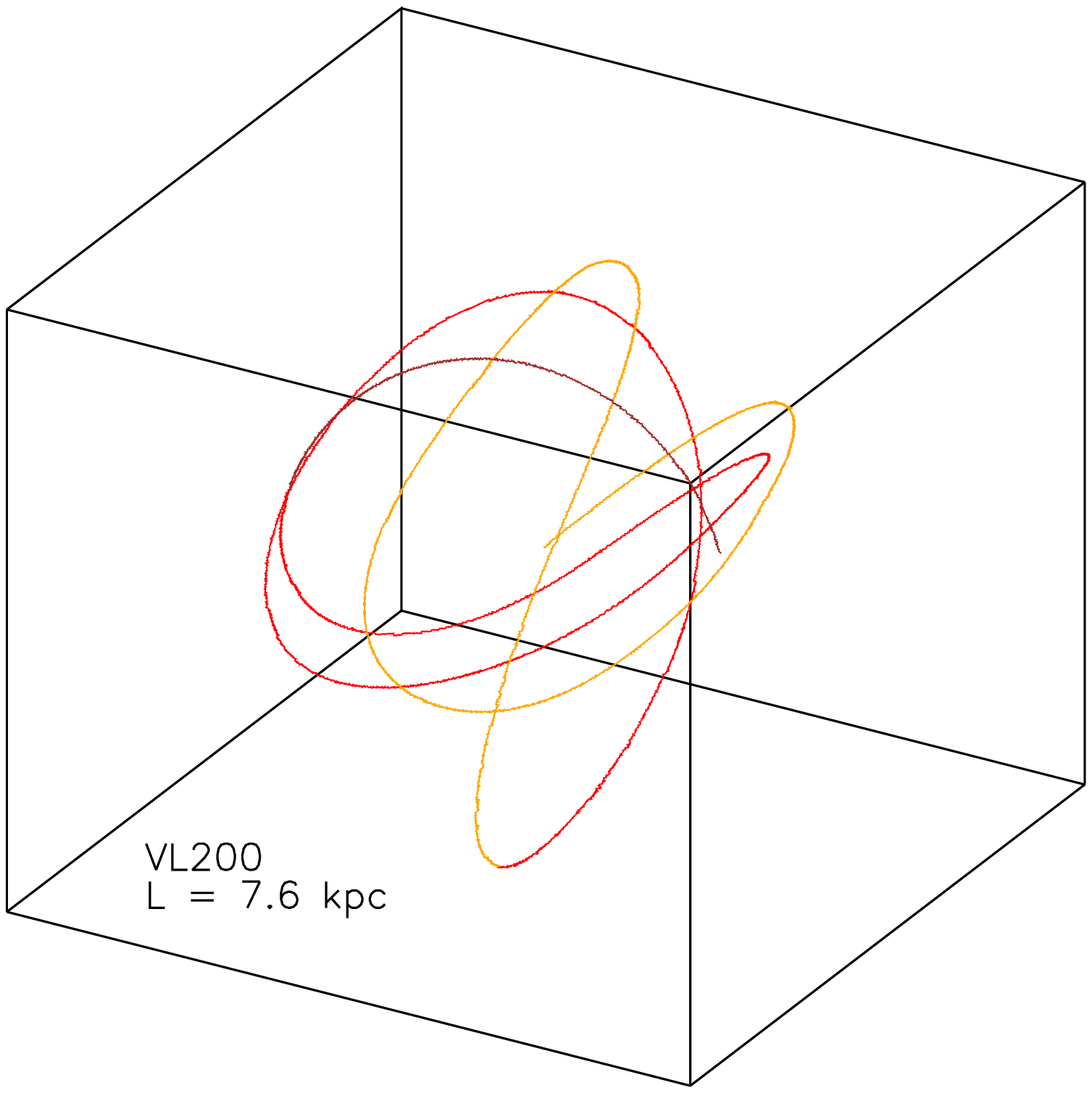}
\includegraphics[width=0.45\textwidth]{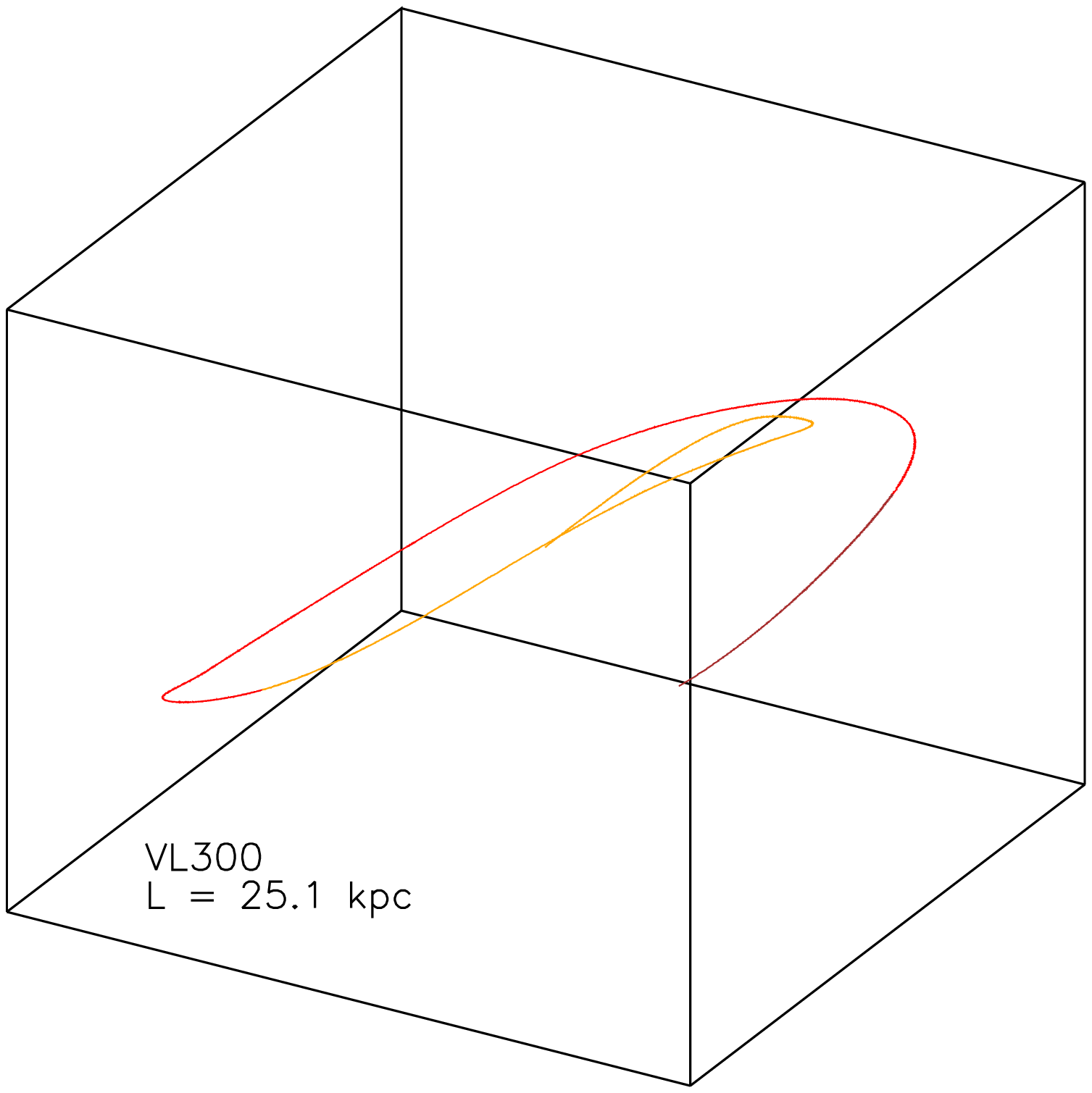}
\caption{\footnotesize {\it Left top, left bottom, and right bottom panels:} Three-dimensional orbits of the recoiling MBHs 
in the VL120, VL200, and VL300 simulations. The first 0.5 Gyr are plotted in yellow, the following 0.5 Gyr 
in red, and the remaining 0.1 Gyr in purple. Box sizes are 2.5, 7.6, and 25.1 kpc, respectively. {\it Right top panel:} 
Comparison between orbits in VL080 ({\it yellow}) and the corresponding energy-conserving orbits in the massless hole 
simulation ({\it green}).}
\label{orb3d}
\vspace{0.3cm}
\end{figure}

The resulting trajectories are shown in Figure~\ref{all_runs}, the orbits' parameters are listed 
in Table 2, and the three-dimensional rendering of the orbits in simulations VL120, VL200, and VL300 
are shown in Figure~\ref{orb3d}. Only $V_{\rm kick}>300\,\kms$ trajectories actually sample the outer 
halo with pericenter distances $R_{\rm max}\gta 30$ kpc, and only $V_{\rm kick}<120\,\kms$ trajectories 
return within 0.5 kpc from the center during the duration of the simulation. The motion of the hole remains 
nearly rectilinear for one or two oscillations only, as the $y$- and $z$-components  
of its orbit become rapidly important due to asphericities in the halo potential. This increases 
the MBH decay timescale compared to a spherical model, as we show below. 
Dynamical friction has only a weak effect on the maximum displacement of the MBH. This can be seen in 
Figure~\ref{orb3d} (right top panel), where a sixth simulation was carried out with the recoiling 
hole treated as a massless test particle of initial kick velocity $V_\mathrm{kick}=80\,\kms$.
A comparison with VL080 shows how, for the first 2-3 oscillations, dynamical friction does not 
strongly influence the motion of the hole, and the maximum displacement is similar to that of the
energy-conserving orbit. It is only at later times that the effect of friction sets in, reducing the amplitude and 
period of the oscillations and bringing the hole back to the center. Note how, for $V_{\rm kick} \geq 120\,\kms$, 
and because of the aspheric nature of the halo, the MBH spends most of its time $>0.8$ kpc away from the center 
and does not have a significant dynamical heating effect on the dark matter distribution in the nucleus. 

\subsection{Orbits in a spherical NFW halo}\label{sph}

It is interesting at this stage to compare the results of our numerical simulations with a semi-analytic 
model of the motion of a recoiling MBH in an NFW halo.  Such a model will allow us to follow the 
trajectory of a recoiling black hole for a Hubble time or until it returns to the center. We define the 
return time, $t_{\rm return}$, as the time it takes for the MBH to decay to within $r=1$ pc of the 
center of the halo with $|E/E_{\rm in}| < 0.001$, where $E$ is the total energy (kinetic + potential) 
of the MBH and $E_{\rm in}$ is its initial energy.  The energy condition is set to ensure that the MBH 
is not simply going through a close periastron passage.

We start by approximating the potential as spherically symmetric and static, with the
$z=0$ host halo parameters given in Table 1. Under these assumptions the trajectory is 
purely radial, and the damping force from the background dark matter can be approximated by the 
classical Chandrasekhar dynamical friction formula \citep{chandrasekhar43,binney87}. The 
corresponding equation of motion is
\begin{equation}\label{classical}
\frac{d\vec{v}}{dt} = - \frac{GM(<r)}{r^3}{\vec{ r}}  - \frac{4 \pi 
G^2 \ln \Lambda \rho(r) M_\mathrm{BH}}{v^3} \left[{\rm erf} (X)-\frac{2X}{\sqrt{\pi}} 
e^{-X^2} \right ] \vec{v},
\end{equation}
where $X \equiv v/\sqrt{2}\sigma$.

The proper definition of the Coulomb logarithm, $\ln \Lambda$, has been extensively 
debated. It is generally defined as $\ln \left (b_\mathrm{max}/b_\mathrm{min}\right)$, 
where the maximum impact parameter $b_\mathrm{max}$ is the scale radius $R_s$ of 
the dark matter distribution, and the minimum impact parameter $b_\mathrm{min}$ is 
the radius of influence of the MBH, $R_{\rm BH}=GM_{\rm BH}/\sigma^2$. Several studies 
\citep[e.g.,][]{colpi99,hashimoto03}  have shown that a dynamically varying value for $\ln \Lambda$  
provides a better estimate of dynamical friction than a constant value when compared to $N$-body 
simulations. Here we follow the treatment of \cite{maoz93}, and in the approximation 
of a spherical NFW host write the Coulomb logarithm as
\begin{equation}
\ln \Lambda \rightarrow \int_{d}^{R_s}\frac{\rho(r)}{\rho_0 r}\,dr = \frac{\rho_s}{\rho_0}\int_{x_0}^{1}\frac{\,dx}{x^2 (1+x)^2} = \frac{\rho_s}{\rho_0} \left [2 \ln \left (\frac{x+1}{x} \right ) -\frac{2x+1}{x(x+1)} \right ]^{1}_{x_0},
\label{log}
\end{equation}
where $\rho_0$ is the central mass density, $x=r/R_s$, $d$ can be interpreted as the 
minimum impact parameter of the Chandrasekhar formula. Throughout this paper we use $\rho_0 = \rho(r={\rm 20\,pc})$ and set $d=R_{\rm BH}$, the radius of influence of the MBH.

\begin{figure}[thb]
\centering
\includegraphics[width=0.9\textwidth]{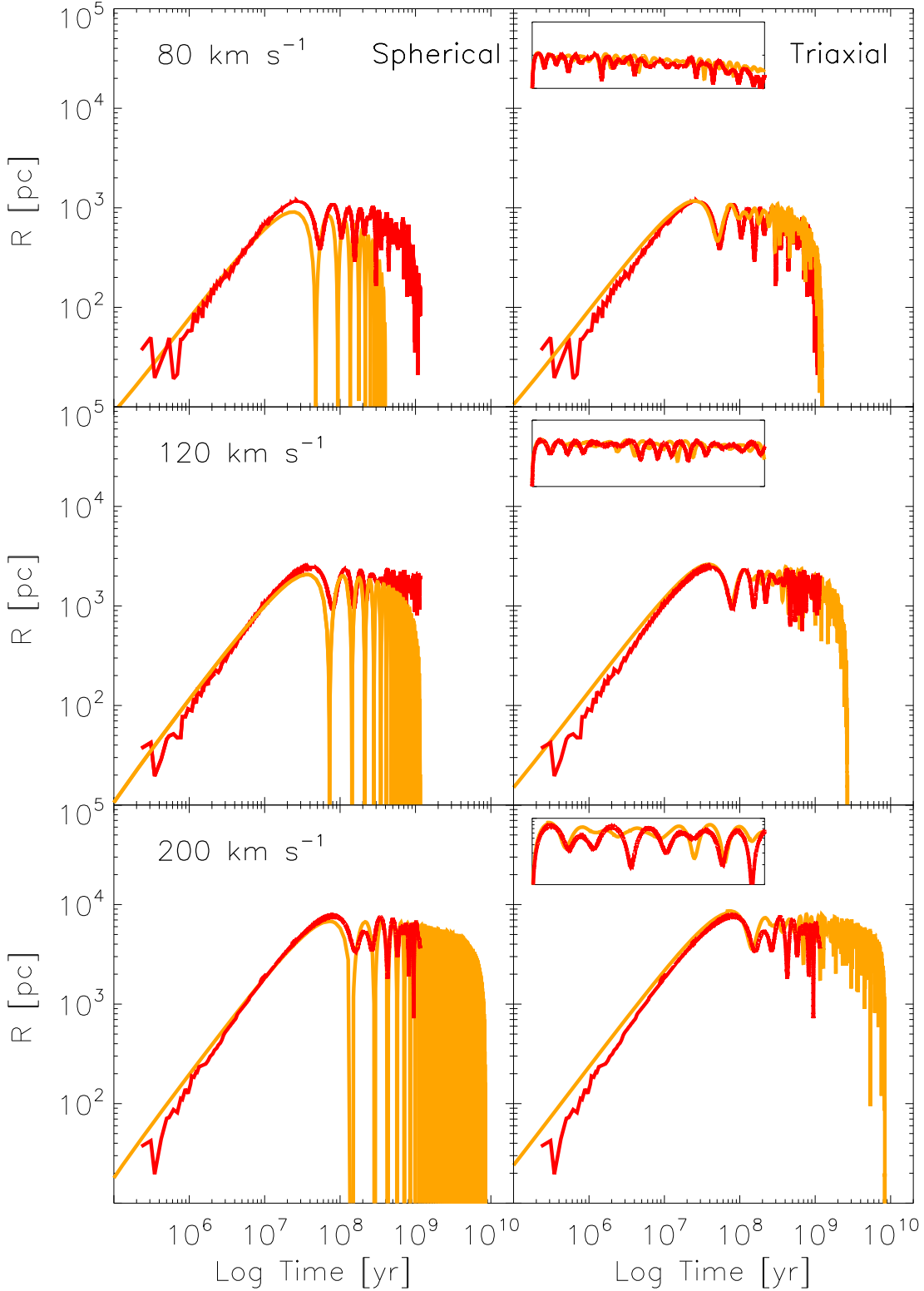}
\caption{Decay histories of recoling MBHs. The radial distance $R$ from the center is plotted as a function of time 
for a spherical NFW ({\it left panel}) and a triaxial NFW halo ({\it right panel}). The $N$-body simulation results  
({\it red curves}) are superposed to the semi-analytic orbit integrations according to model A ({\it orange}). The insets are a close-up of the respective orbit over a timescale of 1.1 Gyr.}
\label{orbits_both}
\end{figure}

The one-dimensional velocity dispersion $\sigma$ for an NFW profile can be solved numerically from the Jeans equation or approximated analytically for $x=r/R_s$ between 0.01 and 100 by the function \citep{zentner03} 
\begin{eqnarray}\label{sigma_nfw}
\sigma^2(x) &=&  V_{200}^2 \frac{c}{f(c)} x (1 + x^2) \int_x^{\infty} \frac{f(x')}{x'^3 (1+x')^2} \,dx' \\
            &\simeq& V_\mathrm{max}^2 \left (\frac{1.4393 x^{0.354}}{1+1.1756 x^{0.725}} \right)^2.
\end{eqnarray}
 We integrate the equation of motion 
numerically using an adaptive Adams-Bashforth-Moulton integration scheme. The resulting radial orbits 
for kick velocities $V_\mathrm{kick}=80,120$, and $200\,\kms$ are shown in Figure~\ref{orbits_both} 
(left panels).  The decay timescale of a recoiling hole in a spherical potential is significantly 
shorter  compared to the results of $N$-body simulations, mainly because of the efficiency 
of dynamical friction during each passage through the nuclear regions. In 
the $V_{\rm kick}=120\,\kms$ case, for example, the MBH is back to the center after 0.6 Gyr 
in the spherical case, while it is still wandering close to $R_{\rm max}$ in the simulations.
A self-consistent estimate of the decay timescale must include the flattening of the cuspy central density profile by 
the oscillating hole. Such a cumulative heating effect, however, is negligible in this case, 
since due to the triaxiality of the potential the MBH does not affect the central density and 
velocity dispersion profiles dramatically. 

\subsection{Orbits in a triaxial NFW halo}\label{tri_model}

The next-order approximation is to model the motion of the recoiling hole in a triaxial, 
dynamically evolving NFW dark matter halo, using the VL-I halo parameters given in Table 
1 and the potential shape parameters plotted in Figure \ref{shapes}. The orbit of the 
hole is fully specified by the conservative force of the dark matter potential $\nabla\Phi$ 
and the damping frictional term:
\begin{equation}
\frac{d\vec{v}}{dt} = -\vec{\nabla}\Phi + \vec{ f}_\mathrm{DF}, 
\end{equation}
where 
\begin{equation}
\Phi= -\frac{G M_{200}}{f(c)}\frac{\ln(1+r_e/R_s)}{r_e},
\end{equation}
and
\begin{equation}
r_e \equiv \left ( x^2 + \frac{y^2}{q^2} + \frac{z^2}{s^2}\right )^{1/2}
\end{equation}
is the ellipsoidal radius. Here $q$ and $s$ are the time- and radial-dependent axis ratios 
defined in Section 2, and $x,y,z$ are Cartesian coordinates along the principal axis of the potential ellipsoid. 
Equation (\ref{classical}) is no longer valid in a triaxial system, where the 
velocity dispersion is non-isotropic and the velocity distribution deviates 
from Maxwellian. We adopt the \cite{pesce92} generalization of the dynamical 
friction formula to triaxial systems \citep[see also][]{vicari07}, 
\begin{equation}
\vec{f}_{\rm DF} = -\Gamma_a V_a \hat{e}_a  -\Gamma_b V_b \hat{e}_b  
-\Gamma_c V_c \hat{e}_c,
\label{TDF}
\end{equation}
\noindent where $V_i$ are the components of the black hole velocity 
along the principal axes $\hat{e}_i  = \{\hat{e}_a,\hat{e}_b,\hat{e}_c\} $ of the local 
velocity dispersion ellipsoid with $a  > b > c$, and $\Gamma_i$ are the dynamical friction coefficients. 
These are given by 
\begin{equation}
\Gamma_i = \frac{2\sqrt{2\pi}G^2 \rho(r_e) \ln \Lambda (M_\mathrm{BH})}{\sigma_1^3} \times B_i(\vec{V},{\bf \sigma}),
\label{tdf} 
\end{equation}
where the velocity dispersion integral is given by
\begin{equation}\label{eqnB_i}
B_i = \int_0^{\infty} \frac{\exp(-\sum_{i=1}^3 
\frac{V_i^2/ 2\sigma_i^2}{\epsilon_i^2 + u})}{\sqrt{(\epsilon_1^2 +u)
(\epsilon_2^2+u)(\epsilon_3^2+u)}} \frac{1}{\epsilon_i^2+u} \,du,
\end{equation}
$\epsilon_i \equiv \sigma_i/\sigma_1$, $\sigma_i^2  = \{\sigma_a^2,\sigma_b^2,\sigma_c^2\}$ 
is the velocity dispersion along the direction $\{\hat{e}_a,\hat{e}_b,\hat{e}_c\} $,  $\sigma_1$ 
is the largest eigenvalue, and $\rho(r_e)$ is the local mass density at the MBH's elliptical radius. In order to calculate the triaxial density 
profile, we deform the spherical density contours in such a way that the volume is preserved. In this approximation, the characteristic 
elliptical radius of the halo  becomes $R_{e,200} = (q\,s)^{-1/3}R_{200}$.

\begin{figure}[thb]
\centering
\includegraphics[width=.45\textwidth]{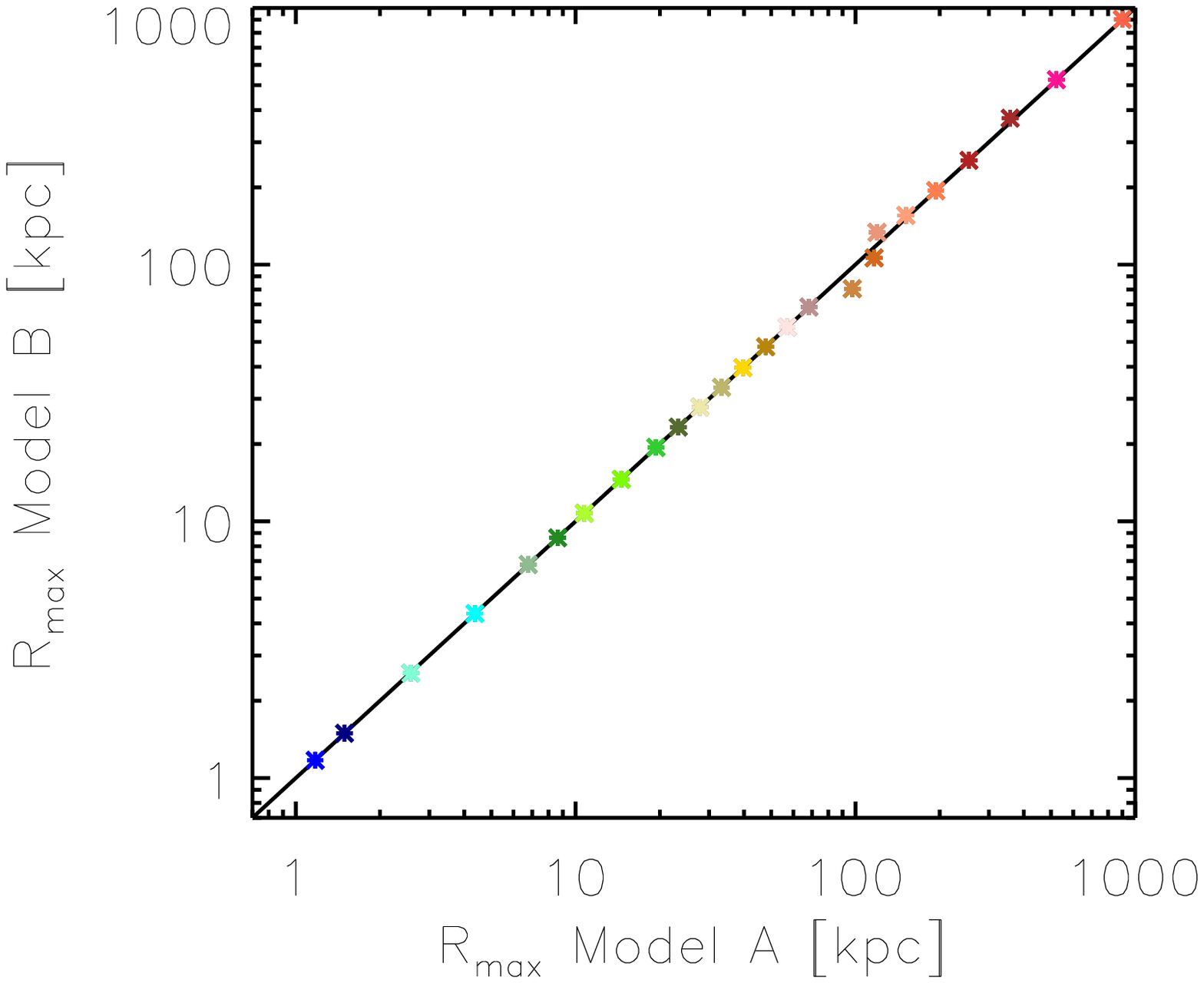}
\includegraphics[width=.45\textwidth]{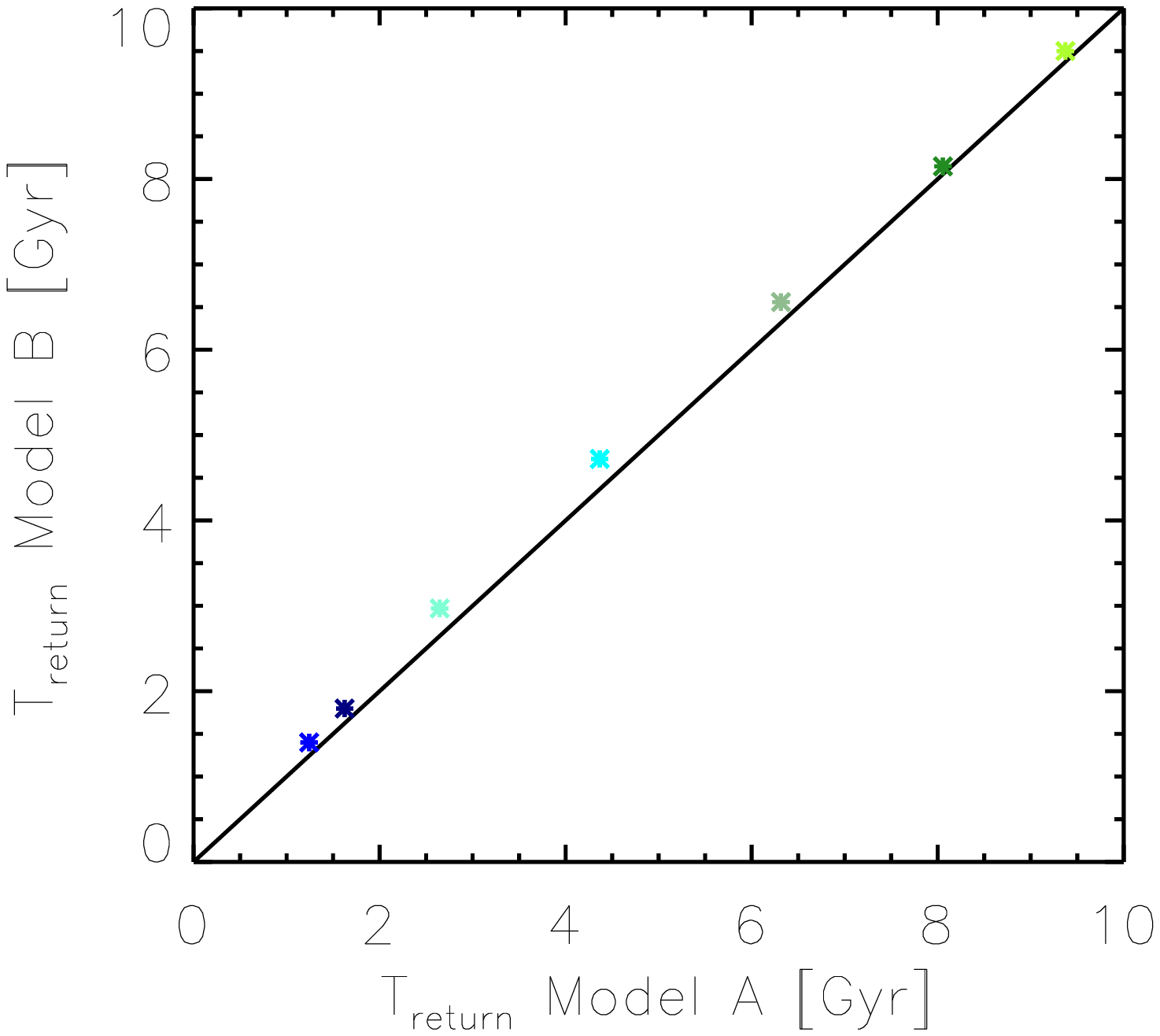}
\caption{{\it Left:} Maximum displacement distance in model A (fiducial) compared to model B for kick velocities in 
the range $80< V_{\mathrm{kick}}<600\,\kms$ ({\it asterisks}). {\it Right:} Return times of models A and B for kick 
velocities $80< V_{\mathrm{kick}}<250\,\kms$ ({\it asterisks}).  Colors represent magnitude of the recoil velocity from $V_{\rm kick}=80\,\kms$ ({\it blue}) to $V_{\rm kick}=600\,\kms$ ({\it red}). }
\label{modelAB}
\end{figure}

A correct estimate of the velocity dispersion as a function of radius and redshift is crucial in the calculation of dynamical 
friction. Here, we take the following approach. First, we measure the ``true" shape and orientation of the local velocity 
dispersion ellipsoids directly from the VL-I simulation (model A; see Section ~\ref{local_quantities} for details). Next, 
we construct a model to calculate the velocity dispersion from the Jeans equation (model B; see Section~\ref{modelB}). 
In model B we neglect streaming motions and assume that the local velocity dispersion ellipsoids are aligned with the 
global potential shape, which results in an overestimate of the velocity dispersion integral $B_i$ given by 
Equation~\ref{eqnB_i}. To normalize model B to the fiducial model A, we introduce a linear fitting factor 
$\eta$ where $B_i^{\rm model\,A} = \eta B_i^{\rm model\,B}$. The main characteristics of the MBH orbits are well 
reproduced by models A and B for a large range of recoil velocities using $\eta=0.5$ (see Figure~\ref{modelAB}).

The resulting orbits are shown in the right panels of Figure~\ref{orbits_both}. The triaxial halo model qualitatively 
reproduces the results of the simulations, the highly non-radial MBH trajectories, and the extended wandering 
times of kicked holes.  Return timescales exceed 10 Gyr already for $V_{\rm kick}=200\,\kms$ (see the last column of Table 2).

\subsubsection{Model A: Local Velocity Dispersions Measured in VL-I}\label{local_quantities}

The local properties of the halo relevant to the calculation of dynamical friction were measured 
from the VL-I simulation as a function of redshift for 10 snapshots in the range $0<z <1.54$, following the method of 
\cite{zemp09}. At each redshift seven distances $r = 1,8,25,50,100,200,400$ kpc from the halo center were randomly 
sampled with 10 spheres of radius
\begin{equation}
r_{\mathrm{sph}}(r) = r_{\mathrm{sph}}(8\,\mathrm{kpc}) \left [ \frac{\rho(8\,\mathrm{kpc})}{\rho(r)} \right ]^{1/3},
\end{equation}
where $r_\mathrm{sph}(8\,\mathrm{kpc}) = 0.5$ kpc and $\rho(r)$ is the spherically averaged mass density 
at radius $r$. In each sphere we measure the local density and calculate the six components of the 
symmetric velocity dispersion tensor, $\sigma_{ij}^2 \equiv \langle v_i v_j\rangle - 
\langle v_i\rangle \langle v_j \rangle$ (here the indices $i$ and $j$ indicate the components along 
the principal axes of the global potential ellipsoid). We then diagonalize the dispersion tensor to obtain a set of eigenvalues and 
eigenvectors. The eigenvalues, $\{\sigma_a^2,\sigma_b^2,\sigma_c^2\}$, are the components of the velocity dispersion in the 
$\hat{e}_i$ basis. 

\begin{figure}[thb]
\centering
\includegraphics[width=0.45\textwidth]{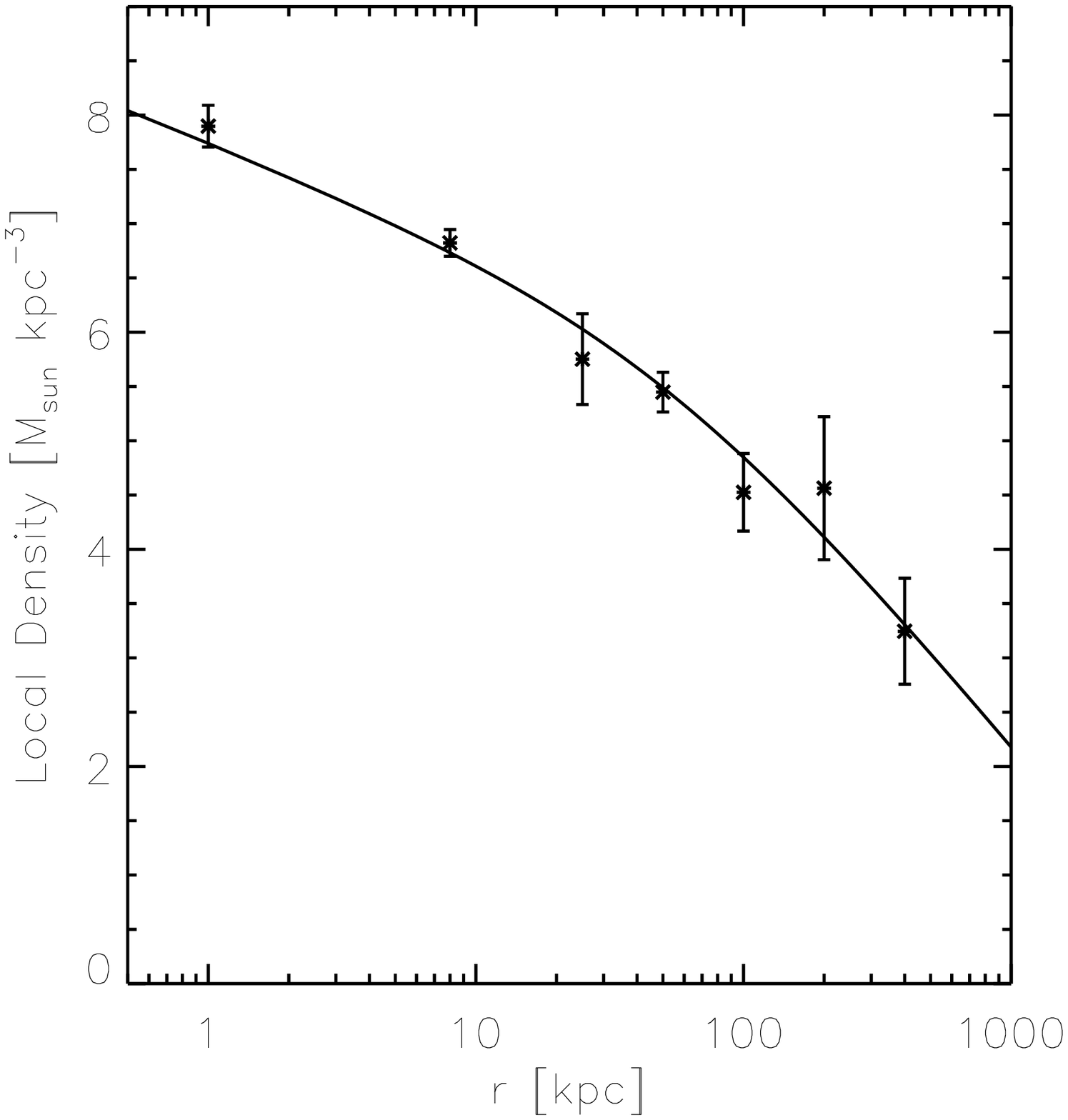}
\includegraphics[width=0.45\textwidth]{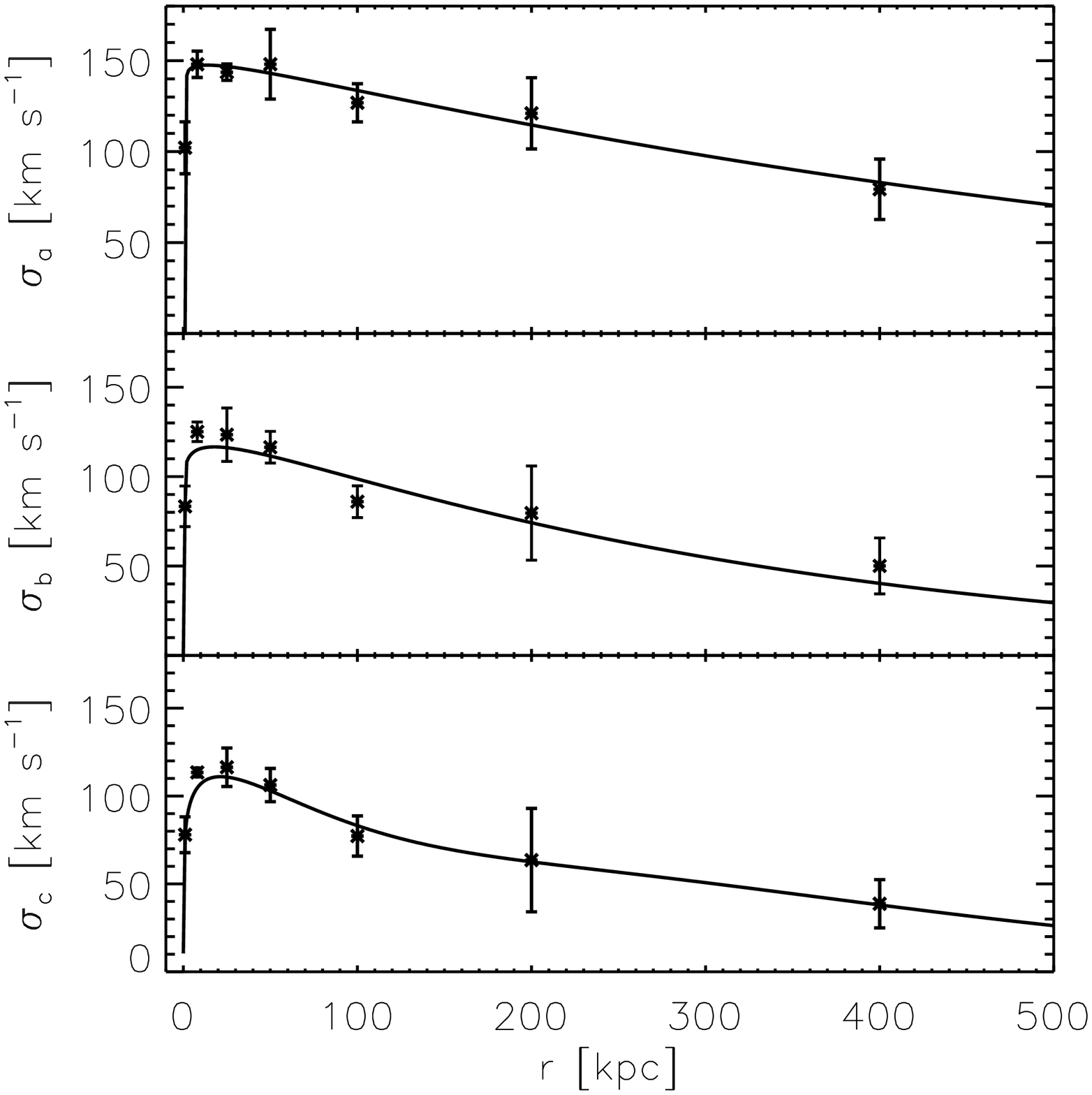}

\caption{Properties of the VL-I main halo at $z=0$. {\it Left:} Local density averaged over an ensemble of spheres at discrete radii from the VL-I halo 
center ({\it asterisks}) and best fit NFW profile ({\it solid line}). {\it Right:} Average local velocity 
dispersion along the principal axes of the local velocity dispersion ellipsoids as a function of radius 
({\it asterisks}) and best fit velocity dispersion profile ({\it solid line}). The error bars represent the dispersion 
around the mean value.}
\label{local_dens}
\end{figure}

For computational convenience, we fit an analytical function to the mean value of the local velocity 
dispersion in all spheres at each radii. This function has the form \citep{pesce92} for model A:
\begin{equation}
\sigma_{i,A}^2(r) = A_i \left [ \frac{1+B_ir^{m_i}}{1+D_i r^{n_i}} \right ] e^{-r/{C_i}},
\end{equation}
where $A_i$ through $D_i$ and $m_i,n_i$ are the best fit values to the velocity dispersion profile 
in the $i$th direction at a given redshift. The parameters at $z=0$ are given in Table~\ref{disp_params},
and the corresponding best-fit curves for $\sigma_{i,A}$ are shown in Figure~\ref{local_dens}b.

The orientation of the local velocity ellipsoids with respect to the global shape was also measured as a 
function of radius and redshift. Table~\ref{spheres_summary} shows the angles between the major, medium, and minor axes of the velocity dispersion ellipsoid and their counterparts in the global potential ellipsoid ($\bar{\alpha},\bar{\beta},\bar{\gamma} $ respectively) averaged over the ensemble of spheres. 
The principal axes of the velocity ellipsoid show significant misalignment 
with the principal axes of the global potential shape: the distribution of orientation angles is 
quite isotropic and cannot be fit by a simple function. 
In our fiducial semi-analytical model (model A), the 
orientation of the local velocity dispersion is obtained by interpolating a grid of mean orientation angles 
as a function of position and redshift at each time step of the numerical integration. Then a random value is drawn in 
the range allowed by the dispersion associated with the mean.

The local density profile is shown in Figure~\ref{local_dens}a. The points represent the average local 
measurement, $\bar{\rho}$, and the error bars are the dispersion around $\bar{\rho}$, labeled 
$\sigma(\bar{\rho})$ in Table~\ref{spheres_summary}. The solid line represents the best fit NFW profile 
to the local density average at $z=0$ (see Table~\ref{table_halo}). 

\subsubsection{Model B: A simple treatment of local velocity dispersion}\label{modelB}
While our fiducial model accurately reproduces important features of the orbits of MBHs in a triaxial potential, 
having a simple prescription to calculate the velocity dispersion analytically would allow us to generalize 
our model and include the effect of other galactic components (see below). 
In this toy model, we assume that the local velocity dispersion ellipsoids are aligned with the potential shape: 
therefore $\{\hat{e}_a,\hat{e}_b,\hat{e}_c\} = \{\hat{e}_x,\hat{e}_y,\hat{e}_z\}$ and all off-diagonal terms of the local 
velocity tensor vanish. We further assume that the halo is in steady state at each snapshot and that there are no streaming motions. 
Under these assumptions we solve for the velocity dispersion along the $i$th coordinate from a simplified Jeans equation:
\begin{equation}\label{sigmai}
\sigma^2_{i,B} = \frac{1}{\rho_e} \int_{x_i}^{\infty}\rho(r_e) \frac{\partial \Phi(r_e)}{\partial x'_i} \,dx'_i,
\end{equation}
where $\rho_e$ is the density at the elliptical radius corresponding to the position of the MBH. We normalize 
the velocity dispersion integral (Equation~\ref{eqnB_i}) to $\eta B_i(\vec{V},\vec{\sigma_B})$ in order to 
match the results of model A. Figure~\ref{modelAB} shows a comparison of models A and B: maximum displacement 
distance and return times are accurately 
reproduced by model B for a large range of kick velocities with $\eta=0.5$.  This analytical representation of the velocity 
dispersion in a triaxial potential proves useful in the construction of the composite potential described below.

\subsection{Orbits in a triaxial NFW halo plus a stellar bulge}\label{bul_model}

A realistic study of the trajectories of recoiling holes must include the gravitational and frictional effect 
of a stellar bulge. Our final set of semi-analytic orbit integrations uses a two-component galaxy model 
consisting of a time-varying triaxial halo (with same parameters as above) and a fixed spherical bulge 
of stellar density
\begin{equation}
\rho_*(r)=\frac{\sigma^2_*}{2 \pi G (r^2 + r_c^2)},
\end{equation}
with isotropic stellar velocity dispersion $\sigma_* = 75\,\kms$, suitable for a Milky-Way-sized host. In the inner regions of the bulge, where stars are the dominant source of dynamical friction, the sphere of influence of the black hole is given by $R_{\rm BH}=GM_{\rm BH}/\sigma^2_*$. The stars within this radius are bound to the black hole and do not exert dynamical 
friction, and therefore a MBH traveling through the very center of the bulge will experience an effective core 
radius $r_c = R_{\rm BH}$. We truncate the bulge profile at an outer radius of $r_{\rm b}=3$ kpc in order to obtain a 
finite bulge mass at large radii, where the dark matter halo dominates the potential. In this model, the mass of the stellar 
bulge within the outer truncation radius is $M_*(<r_b) = 8 \times 10^9 M_{\odot}$.

To find the velocity dispersion tensor of the composite profile, $\sigma^2_{ij}$, we solve the Jeans equations 
under the assumption that the velocity ellipsoid is aligned with the axes of the (dynamically evolving) triaxial 
NFW potential. Thus, the three principal components of the velocity dispersion tensor are given by 
\begin{equation}
\sigma^2_{i,\rm tot} = \frac{1}{\rho_{\rm tot}} \int_{x_i}^{\infty}\rho_{\rm tot} \frac{\partial \Phi_{\rm tot}}{\partial x_i} \,dx_i
\label{sigmatot}
\end{equation}
where $\rho_{\rm tot}$ and $\Phi_{\rm tot}$ are the total (NFW halo + stellar bulge) density and potential. We calculate the
Coulomb logarithm from Equation~\ref{log} using the total mass density and $d=GM_{\rm BH}/\sigma^2$, 
where the composite velocity dispersion is now given by $\sigma_{\rm tot} = \sqrt{\sigma_a^2+\sigma_b^2+\sigma_c^2}$ 
(with $\sigma_i$ given by Equation \ref{sigmatot}). As in Section~\ref{modelB}, we normalize the velocity dispersion integral to $\eta B_i(\vec{V},\vec{\sigma}_{\rm tot})$, where $B_i(\vec{V},\vec{\sigma}_{\rm tot})$ is the velocity dispersion integral of the composite potential. We assume that the spherical stellar bulge fully dominates the potential in the region $r< 100$ pc and therefore dynamical friction is well approximated by the Chandrasekhar formula. We fit for $\eta$  by comparing orbits obtained 
with Equation~\ref{classical} with those obtained with Equation~\ref{TDF} for $V_{\rm kick} < 250\,\kms$. The best fit yields
$\eta=0.1$. 

\begin{figure}
\begin{center}
\includegraphics[scale=0.7]{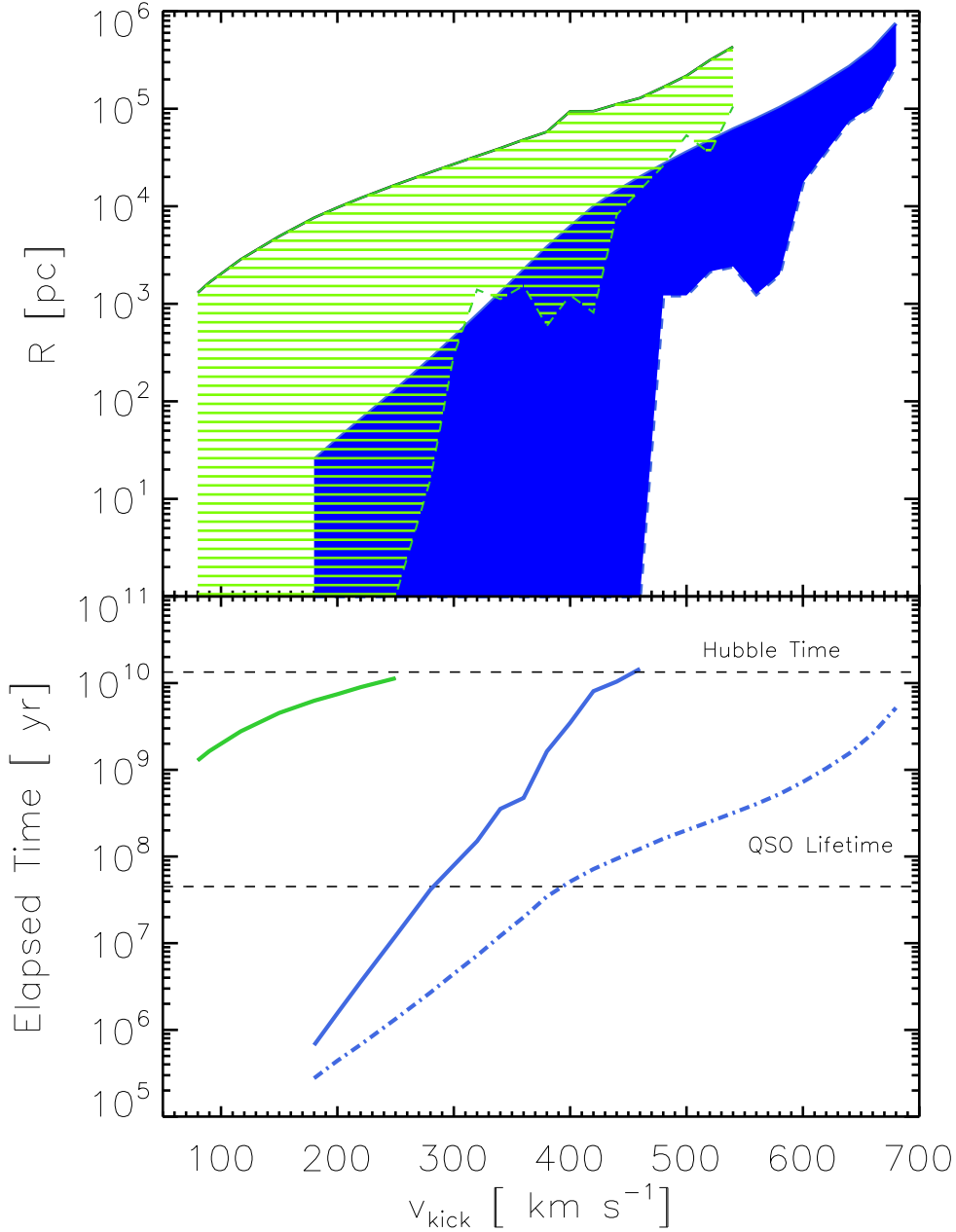}
\caption{{\it Upper panel:} A set of apocenter distances ({\it solid line}) and pericenter distances ({\it dashed line}) 
for a recoiling MBH of mass $M_{\bullet} = 3\times 10^6\,\msun$ in a triaxial Milky Way-sized dark matter only host 
({\it green}) and dark matter + bulge host ({\it blue}). The colored areas show the corresponding regions in the 
$R-V_{\rm kick}$ plane occupied by the wandering holes. {\it Lower panel: } Return timescales of a MBH in a dark matter 
only host ({\it green line}) and a dark matter + bulge potential ({\it solid blue line}). Also shown is the time it takes to the
hole to reach its apocenter ({\it dashed-dotted blue line}).}
\label{maxs}
\end{center}
\end{figure}

Table 4 gives the MBH apocenter, its pericenter, and the return time calculated using our two-component 
model: (1) for  $V_{\rm kick}>460$ $\kms$ we stopped numerical integration after a Hubble 
time $t_{\rm H}$, while the hole was still wandering  tens to hundreds of kiloparsecs away from the center; 
(2) for kick velocities below  $380\,\kms$,  dynamical friction against bulge stars now efficiently 
damp the motion of the MBH already on the first outward trajectory, and reduces the decay timescale 
to less than 2 Gyr. Recoiling holes do not leave the bulge; (3) for the maximum kick velocities 
predicted in the case of non-rotating holes, $V_{\rm kick}\lta 200\,\kms$, the MBH reaches a maximum
distance of only 40 pc from the center and decays back within 2 Myr; (4) black holes that leave the 
stellar bulge and enter the triaxial dark matter halo do not return to the center within a Hubble time. 
The pericenter distances, apocenter distances, and the return times of MBHs are shown in Figure~\ref{maxs} for a dark matter only 
potential and a more realistic dark matter + bulge potential. According to the latter model, a MBH which is kicked with initial velocity 
$V_{\rm kick}=400$ $\kms$ reaches $R_{\rm max}$ before $10^8$ yr, a time comparable with the typical QSO lifetime, and spends 
most of its time orbiting at a distance $r >1$ kpc away from the center of the bulge.
\section{Summary}

Coalescing MBH pairs will give origin to the loudest gravitational wave events in
the universe, and are one of the primary targets for the planned {\it Laser
Interferometer Space Antenna} ({\it LISA}; e.g., Sesana \etal 2004). 
The anisotropic emission of gravitational waves also removes net linear momentum from the binary and
imparts a kick to the center of mass of the system. The outcome of this ``gravitational rocket'' 
has been the subject of many recent numerical relativity studies. Non-spinning holes
recoil with velocities below 200 $\kms$ that only depend on the binary mass ratio, while much larger kicks require rapidly rotating holes. 
Little is known about the masses of 
MBH binaries and their spins: the distribution of all binary mass ratios expected in some
hierarchical models of the co-evolution of MBHs and their hosts is found to be relatively 
flat \citep{volonteri08}: if it is not ejected from the host 
altogether, the recoiling MBH will travel some maximum distance and then return towards 
the center on a decay timescale that depends on the shape of the potential and on 
the effectiveness of  gas drag and dynamical friction against the stars and the dark matter of the host galaxy.

We have carried out a detailed study of the fate of bound recoling
holes in Milky Way-sized potentials, running  $N$-body simulations of the motion of a  
$M_{\rm BH}=3.7\times 10^6\,\msun$ MBH remnant in the ``Via Lactea I''  dark matter halo.
In the simulations, the MBH receives  a kick velocity of $V_{\rm kick}$ = 80, 120, 200, 300, and 
400 $\kms$ following the coalescence of its progenitor binary, and moves within the ``live" host 
subject only to gravity and dynamical friction against the dark matter background. We have
used these calculations to build realistic semi-analytic models of the hole's trajectory in a 
time-varying triaxial NFW potential, where the dynamical friction force is calculated directly from the velocity 
dispersion tensor, and in a two-component  triaxial halo+spherical bulge model. The latter case 
should offer a more realistic picture of the dynamics of kicked MBHs in situations where gas drag, friction by disk
stars, and the heating effect of the returning hole on the central cusp are all negligible. Our results on 
the trajectories of recoiling MBHs can be summarized as follows: 

\begin{enumerate}

\item  Owing to asphericities in the  dark matter potential, the black hole's orbits are highly non-radial, resulting 
in a significantly increased decay timescale compared to the spherical case. This is in qualitative 
agreement with earlier results by \citet{vicari07}.

\item In a triaxial NFW halo return timescales to the center exceed 5 Gyr already for $V_{\rm kick}=200\,\kms$, 
and are longer than the Hubble time for $V_{\rm kick}\ge 250\,\kms$.

\item In a triaxial halo+spherical bulge potential, decay timescales are much shorter than in the bulgeless 
case. For kick velocities  $V_{\rm kick}<380\,\kms$,  dynamical friction against bulge stars now efficiently 
damp the motion of the MBH already on the first outward trajectory, and reduces the decay timescale 
to less than 2 Gyr. For recoil velocities $V_{\rm kick}>500\,\kms$ the MBH does not return to the center of 
its host within a Hubble time. Recoling black holes do not leave the bulge and remain within a few 
tens of parsecs from the center for $V_{\rm kick}\lta 200\,\kms$.

\end{enumerate}

A kicked MBH  can retain the inner parts of its accretion disk, providing fuel for a continuing 
luminous phase along its trajectory. Let us assume all recoiling holes
accrete at a fraction $f_E$ of the Eddington rate $\dot M_E=4\pi G M_{\rm BH}m_p/(c\sigma_T
\epsilon)$, where $\epsilon$ is the radiative efficiency. The duration of the luminous phase
depends on the amount of disk material out to the radius $R_{\rm out}\approx
{GM_{\rm BH}/V_{\rm kick}^2}$ that is carried by the hole. In the case of an
$\alpha$-disk, this is given by  (Loeb 2007)
\begin{equation}
M_{\rm disk}\approx (1.9\times 10^6\,\msun)~\alpha_{-1}^{-0.8} (\epsilon_{-1}
/f_E)^{-0.6} M_{7}^{2.2} V_{3}^{-2.8},
\label{mmax}
\end{equation}
where $\epsilon_{-1}\equiv \epsilon/0.1$, $M_{7}\equiv M_{\rm BH}/10^7\,\msun$,
$V_{3}\equiv V_{\rm kick}/10^3\,\kms$, and $\alpha_{-1} \equiv \alpha/0.1$
is the viscosity parameter. The condition $M_{\rm disk}\le M_{\rm BH}$ then requires
\begin{equation}
V_{\rm kick}\ge 550\,\kms ~\alpha_{-1}^{-0.28} (\epsilon_{-1}/f_E)^{-0.21}M_7^{0.43}.
\label{vmax}
\end{equation}
For lower kick velocities $M_{\rm disk}=M_{\rm BH}$, 
corresponding to an AGN lifetime of $t_{\rm QSO}=\epsilon c \sigma_T/(4\pi G
m_p f_E)\approx 4.5\times 10^7\,{\rm yr}\,(\epsilon_{-1}/f_E)$.
A recoiling hole/disk system with ($M_7,\alpha,\epsilon,f_E,M_{\rm disk})=
(1,0.1,0.1,1,M_{\rm BH})$ could then be shining for half a Gigayear as an off-center 
quasar over a large fraction of its ``wandering" phase. 
Thus, cases where the recoil kick is large enough to launch the MBH into the triaxial halo are 
favorable for the detection of off-nuclear quasars. However, if the MBH is initially embedded in a 
gas-rich environment, gas drag may damp its motion significantly \citep{guedes08}, even for moderate kicks, 
lowering the detection probability. Furthermore, the spins of both black holes in a MBH binary tend to align due to torques induced by the surrounding gas, reducing the kick velocity to $v_{\rm kick} < 200\,\kms$ \citep{bogdanovic07}. The motion of a recoiling MBH in a gas-rich merger including a stellar and dark matter component will be the subject of a subsequent paper. 

\acknowledgments
Support for this work was provided by NASA through grant NNX09AJ34G (P.M.) and by NSF 
through a Graduate Research Fellowship (J.G.). M.K. gratefully acknowledges support from the William L. Loughlin
Fellowship at the Institute for Advanced Study. All computations were performed on NASA's Project Columbia
supercomputer system and on UCSC's Pleiades cluster. We thank Joel Primack, Eliot Quataert, Doug Lin, and Lucio Mayer for many useful 
discussions.

\begin{table}
\begin{center}
\caption{VL-I halo parameters}
\begin{tabular}{|c|c|c|c|c|c|c|}
\tableline
$a$ & $\rho_s$ &  $R_s$  & $R_{200}$ & $M_{200}$ & $V_\mathrm{max}$ & $v_\mathrm{esc}$ \\
    & ($10^6$ M$_{\odot}$ kpc$^{-3})$ & (kpc) & (kpc) & ($10^{12}$ M$_{\odot}$) & (km s$^{-1}$) & (km s$^{-1}$) \\
\tableline
0.393 & 0.16 & 38.0 & 194.3 & 1.03 & 160.53 & 488.5 \\
\tableline
0.423 & 0.21 & 36.7 & 213.1 & 1.13 & 163.7 & 498.2 \\
\tableline
0.465 & 0.41 & 31.3 & 233.8 & 1.19 & 167.9 & 510.9 \\
\tableline
0.507 & 0.54 & 30.8 & 250.9 & 1.22 & 166.7 & 507.3 \\
\tableline
0.549 & 0.72 & 29.7 & 271.5 & 1.31 & 170.4 & 518.4 \\
\tableline
0.591 & 0.99 & 28.1 & 292.5 & 1.43 & 176.4 & 536.7 \\
\tableline
0.633 & 1.40 & 26.2 & 311.8 & 1.54 & 182.6 & 555.8 \\
\tableline
0.675 & 1.87 & 25.1 & 327.4 & 1.61 & 186.4 & 567.2 \\
\tableline
0.762 & 2.40 & 26.0 & 356.6 & 1.76 & 189.2 & 575.6 \\
\tableline
0.877 & 3.54 & 26.2 & 376.2 & 1.77 & 187.0 & 569.0 \\
\tableline
0.901 & 3.67 & 26.7 & 379.2 & 1.77 & 185.6 & 564.8 \\
\tableline
0.950 & 4.51 & 26.2 & 384.9 & 1.77 & 185.9 & 565.5 \\
\tableline
1.000 & 5.33 & 25.8 & 389.3 & 1.77 & 185.1 & 566.2 \\
\tableline
\end{tabular}
\end{center}
\vskip 0.3cm
\label{table_halo}
\end{table}

\begin{table}
\label{results_sim}
\centering
\begin{minipage}{0.46\textwidth}
\centering
\caption{N-body Simulation Results}
\begin{tabular}{|c|c|c|c|c|c|c|}
\tableline
Run Name & $V_{\mathrm{kick}}$ & $R_{\mathrm{max}}$ & $R_{\mathrm{min}}$ & $t_{\mathrm{end}}$ & $R_{\mathrm{end}}$ & $t_{\mathrm{return}}$ \\
 & (km s$^{-1}$) & (kpc) & (kpc) & (Gyr) & (kpc) & (Gyr) \\  
\tableline
VL080 & 80 & 1.18 & 0.03 & 1.15 & 0.09 & 1.16 \\
\tableline
VL120 & 120 & 2.49 & 0.56 & 1.15 & 1.90 & 2.78 \\
\tableline
VL200 & 200 & 7.69 & 0.72 & 1.15 & 3.71 & 8.45 \\
\tableline
VL300 & 300 & 28.21 & 1.51 & 1.15 & 14.93 & $> t_{\mathrm{H}}$ \\
\tableline
VL400 & 400 & 83.65 & 22.95 & 1.15 & 22.95 & $> t_{\mathrm{H}}$ \\
\tableline
\end{tabular}
\footnotetext{Columns 2,3,4,5,6 and 7 list the initial kick velocity, the MBH apocenter, its pericenter, 
the end time of the simulation, the distance of the MBH from the halo center at $t_\mathrm{end}$, and 
the return time  calculated using a triaxial NFW model, respectively. The return time, $t_{\rm return}$, 
is defined as the time is takes for the MBH to lose all but 0.1$\%$ of its initial total energy and decay to 
within 1 pc of the center of the halo.}
\end{minipage}
\vskip 0.6cm
\end{table}

\begin{table}
\centering
\begin{minipage}{0.46\textwidth}
\centering
\caption{Best-fit parameters to the velocity dispersion profile at $z=0$.}
\begin{tabular}{|c|c|c|c|c|c|c|}
\tableline
    & A & B & C & D & m & n \\
   & (km$^2$ s$^{-2}$) & (kpc$^{-m}$) & (kpc) & (kpc$^{-n}$)& & \\
 \tableline
$\sigma_a^2$ & $2.24\times10^4$  &  1.145   &   172.21    & 0.0026  &   -$4.87\times10^3$  & 1.1132 \\
$\sigma_b^2$ & $7.16\times10^2$  &  0.567    &  153.55    &  14.300  &   -$1.29\times10^2$ & 0.1217 \\
$\sigma_c^2$ & $3.65\times10^2$ &  0.4102    &  117.02    &  22.698  &   -$0.19\times10^2$ & 0.1646\\
\tableline
\end{tabular}
\label{disp_params}
\end{minipage}
\vskip 0.6cm
\end{table}

\begin{table}
\begin{minipage}{0.9\textwidth}
\caption{Summary of Local Properties at $z=0$}
\begin{tabular}{|c|c|c|c|c|c|c|c|c|}
\tableline
$r$ & (kpc) & 1 & 8 & 25 & 50 & 100 & 200 & 400 \\
\tableline
$\bar{\rho}$ & (M$_{\odot}$ pc$^{-3}$) & 7.90$\times10^{-2}$&   6.65$\times10^{-3}$ & 5.64$\times10^{-4}$ & 2.80 $\times10^{-4} $& 3.35 $\times 10^{-5}$ & 3.65$\times 10^{-5}$ & 1.76 $\times 10^{-6}$ \\
$\bar{\sigma}_a$ & ($\kms$) &  102.1 &    148.0 &      143.7 &     148.1   &   126.9   &   121.1   &  79.31 \\
$\bar{\sigma}_b$ & ($\kms$) & 83.38   &  125.1 &     124.5 &     116.4 &     85.98 &     79.60 &     50.06 \\
$\bar{\sigma}_c$ & ($\kms$) & 77.99 &     113.4 &     116.4 &     106.3 &     77.26 &     63.54 &     38.71 \\
$\bar{\alpha}$ & ($^{\circ}$) & 19.82   &  53.19  &    61.45   &  62.30    &  36.57   &   34.79  &   45.19 \\
$\bar{\beta}$ &  ($^{\circ}$)  & 69.17  &    48.28   &   57.47    &  50.43 &     54.81   &   54.66 &     57.69 \\
$\bar{\gamma}$ & ($^{\circ}$) & 84.10  &    69.04  &    69.31 &    66.32  &    54.63 &    58.78   &   56.22 \\

$\sigma(\bar{\rho})$ & (M$_{\odot}$ pc$^{-3}$) & 5.88$\times10^{-2}$ &  1.67$\times10^{-3}$&  1.16$\times10^{-3}$  & 9.09$\times 10^{-5}$&  3.13$\times 10^{-5}$ & 5.37 $\times 10^{-5}$ &1.26$\times 10^{-6}$ \\
$\sigma(\bar{\sigma}_a)$ & ($\kms$) & 14.27  &   7.253 &      4.468   &   19.15  &   10.48    &  19.58   &   16.61\\
$\sigma(\bar{\sigma}_b)$ & ($\kms$) & 11.36    &  5.470   &   14.94   &   8.877 &    8.874   &   26.34 &    15.64 \\
$\sigma(\bar{\sigma}_c)$ & ($\kms$) & 10.21    &  2.463     & 10.97   &   9.450 &     11.45   &  29.42   &   13.71 \\
$\sigma(\bar{\alpha})$ & ($^{\circ}$) & 32.56    &  37.80  &    29.52    &  25.96 &     24.41   &   14.58    &  15.43 \\
$\sigma(\bar{\beta})$ &  ($^{\circ}$)  & 36.66    &  35.48   &   26.46  &   29.98   &   16.02    & 33.27 &    27.06\\
$\sigma(\bar{\gamma})$ & ($^{\circ}$) & 7.58  &    12.59     & 22.23     & 21.99 &     25.78    & 29.95  &   17.58 \\
\tableline
\end{tabular}
\footnotetext{Halo local properties averaged over an ensemble of 10 spheres at each radius. The rows show the
mass density $\bar{\rho}$, the average velocity dispersion components, ($\bar{\sigma}_a,\bar{\sigma}_b,\bar{\sigma}_c$),
along the principal axes of the velocity dispersion ellipsoid, and the angles, ($\bar\alpha,\bar\beta,\bar\gamma$), 
between the major, intermediate, and minor axes of the local velocity and the global potential ellipsoids. 
Also listed are the dispersions of the above quantities.}
\label{spheres_summary}
\end{minipage}
\vskip 0.1cm
\end{table}

\begin{table}
\label{results_model}
\centering
\begin{minipage}{0.9\textwidth}
\centering
\caption{Semi-analytic Model Results}
\begin{tabular}{|c|c|c|c|}
\tableline
$V_{\mathrm{kick}}$ & $R_{\mathrm{max}}$ & $R_{\mathrm{min}}$ &  $t_{\mathrm{return}}$ \\
 (km s$^{-1}$) & (kpc) & (kpc) & (Gyr) \\  
\tableline
200 & 0.0406 & 0.0010 & 0.0016 \\
\tableline
280 & 0.2707 & 0.0010 & 0.0415 \\
\tableline
300 & 0.4512 & 0.0010 & 0.0791 \\
\tableline
360 & 2.2022 & 0.0010 & 0.4735 \\
\tableline
380 & 3.7714 & 0.0010 & 1.6275 \\
\tableline
400 & 6.8619 & 0.0010 & 3.4846 \\
\tableline
420 & 10.5830 & 0.0010 & 8.0657 \\
\tableline
440 & 17.9090 & 0.0010 & 10.4097 \\
\tableline
460 & 24.0626 & 0.0010 & $> t_{\mathrm{H}}$ \\
\tableline
500 & 37.2263 & 1.2189 & $> t_{\mathrm{H}}$ \\
\tableline
560 & 84.6069 & 1.2555 & $> t_{\mathrm{H}}$ \\
\tableline
600 & 137.3806 & 17.4473 & $> t_{\mathrm{H}}$ \\
\tableline
680 & 786.7245 & 276.3753 & $> t_{\mathrm{H}}$ \\
\tableline

\end{tabular}
\footnotetext{Columns 1,2,3, and 4 list the initial kick velocity, the MBH apocenter, its pericenter, and 
the return time within 1 pc from  the center  calculated using a triaxial NFW + isothermal spherical
bulge model (see the text for details).}
\end{minipage}
\vskip 0.6cm
\end{table}


\begin{thebibliography}{}
\bibitem[Allgood et al.(2006)]{Allgood2006}  Allgood, B., Flores, A.~R., Primack, J.~R., Kravtsov, A.~V., Wechsler, R.~H., Faltenbacher, A., \& Bullock, J.~S. 2006, MNRAS, 367, 1781 
\bibitem[Baker et al.(2008)]{baker08} Baker, J. G., Boggs, W. D., Centrella, J., Kelly, B. J., McWilliams, S. T., Miller, M. C., \& van Meter, J. R. 2008, ApJ, 682, L29
\bibitem[Baker et al.(2006a)]{baker06a} Baker, J. G., Centrella, J., Choi, D.-I., Koppitz, M., \& van Meter, J. 2006a, 
PhysRevL, 96, 111102
\bibitem[Baker et al.(2006b)]{baker06b} Baker, J. G., Centrella, J., Choi, D.-I., Koppitz, M., van Meter, J. R., \& Miller, M. C. 2006b, \apj, 653, L93
\bibitem[Begelman et al.(1980)]{begelman80} Begelman, M.~C., Blandford, R.~D., \& Rees, M.~J. 1980, Nature, 287, 307
\bibitem[Bekenstein(1973)]{bekenstein73} Bekenstein, J. D. 1973, \apj, 183, 657]
\bibitem[Bertschinger(2001)]{bertschinger01} Bertschinger, E.  2001, ApJS, 137,1B
\bibitem[Binney \& Tremaine(1987)]{binney87} Binney, J., \& Tremaine, S. 1987, Galactic Dynamics (Princeton: Princeton Univ. Press)
\bibitem[Blecha \& Loeb(2008)]{blecha08} Blecha, L., \& Loeb, A. 2008, MNRAS, 390, 1311
\bibitem[Bogdanovi\'c et al.(2009)]{bogdanovic09} Bogdanovi\'c, T., Eracleous, M.; Sigurdsson, S. 2009, \apj, 697, 288
\bibitem[Bogdanovi\'c et al.(2007)]{bogdanovic07} Bogdanovi\'c, T., Reynolds, C. S., \& Miller, C. 2007, \apj, 661, L147
\bibitem[Bonning, Shields, \& Salviander(2007)]{bonning07} Bonning, E. W., Shields, G. A., \& Salviander, S. 2007, \apj, 666, L13
\bibitem[Boylan-Kolchin et~al.(2004)]{boylan-kolchin2004} Boylan-Kolchin, M., Ma, C.-P., \& Quataert, E. 2004, ApJ, 613, L37
\bibitem[Campanelli et al.(2006)]{campanelli06} Campanelli, M., Lousto, C. O.,  Marronetti, P., \& Zlochower, Y. 2006, Phys. Rev. Lett., 96, 111101
\bibitem[Campanelli et al.(2007a)]{campanelli07a} Campanelli, M., Lousto C., Zlochower, Y., \& Merrit, D. 2007, \apj, 659, L5
\bibitem[Chandrasekhar(1943)]{chandrasekhar43} Chandrasekhar, S. 1943, \apj, 97, 255
\bibitem[Colpi et al.(1999)]{colpi99} Colpi M. , Mayer L., \& Governato F. 1999, \apj, 525, 720
\bibitem[Diemand et al.(2007a)]{diemand07a} Diemand J., Kuhlen M., \& Madau P. 2007a, \apj, 657, 262
\bibitem[Diemand et al.(2007b)]{diemand07b} Diemand J., Kuhlen M., \& Madau P. 2007b, \apj, 671, 1135
\bibitem[Diemand et al.(2008)]{diemand08} Diemand, J., Kuhlen, M., Madau, P., Zemp, M., Moore, B., Potter, D., \& 
Stadel, J. 2008, Nature, 454, 735
\bibitem[Dotti et al.(2008)]{dotti08} Dotti, M., Montuori, C., Decarli, R., Volonteri, M., Colpi, M., \& Haardt, F. 2008, arXiv:astro-ph/ 0809.3446
\bibitem[Favata et al.(2004)]{favata04} Favata, M., Hughes, S.A., \& Holz, D.E. 2004, \apj, 607, L5
\bibitem[Fitchett \& Detweiler(1984)]{fitchett84} Fitchett, M.J. \& Detweiler, S. 1984, \mnras, 211, 933
\bibitem[Franx et al.(1991)]{franx91} Franx, M., Illingworth, G., \& de Zeeuw, T. 1991, \apj, 383, 112
\bibitem[Ghez et al.(2005)]{ghez05} Ghez, A. M., Salim, S., Hornstein, S. D., Tanner, A., Lu, J. R., Morris, M., 
Becklin E. E., \& Duch\^{e}ne, G. 2005, \apj, 620, 744
\bibitem[Gonz\'alez et al.(2007)]{gonzalez07a} Gonz\'alez, J. A., Hannam, M., Sperhake, U., Br\"{u}gmann, B., \& Husa, S. 2007, Phys. Rev. Lett., 98, 231101
\bibitem[Gualandris \& Merritt(2008)]{gualandris08} Gualandris, A., \& Merritt, D. 2008, \apj, 678, 780 
\bibitem[Guedes et al.(2008)]{guedes08} Guedes, J., Diemand, J., Zemp, M., Kuhlen, M., Madau, P., Mayer, L., \& Stadel, J.  2008, Astron. Naschr., 329, 1004
\bibitem[Hashimoto et al.(2003)]{hashimoto03} Hashimoto, Y., Funato, Y., \& Makino, J. 2003, \apj, 582, 196
\bibitem[Heckman et al.(2009)]{heckman09} Heckman, T., Krolik, J. H., Moran, S. M., Schnittman, J., \& Gezari, S. 2009, \apj, 695, 363
\bibitem[Herrmann et al.(2007)]{herrmann07} Herrmann, F., Hinder, I., Shoemaker, D., \& Laguna, P. 2007, Class. Quantum Grav., 24, 33
\bibitem[Komossa et al.(2003)]{komossa03} Komossa, S., Burwitz, V., Hasinger, G., Predehl, P., Kaastra, J. S., \& Ikebe, Y.  
2003, \apj, 582, L15
\bibitem[Komossa et al.(2008)]{komossa08} Komossa, S., Zhou, \& H., Lu, H. 2008, \apj, 678, 81 
\bibitem[Kormendy et al.(1995)]{kormendy95} Kormendy, J. \&  Richtsone, D. 1995, \araa, 30, 581
\bibitem[Kuhlen et al.(2007)]{kuhlen07} Kuhlen, M., Diemand, J., \& Madau, P. 2007, \apj, 671, 1135
\bibitem[Loeb(2007)]{loeb07} Loeb, A. 2007 Phys. Rev. Lett., 99, 041103
\bibitem[Madau \& Quataert(2004)]{madau04} Madau, P. \& Quataert, E. 2004, \apj, 606, L17
\bibitem[Madau \& Rees(2001)]{madau01} Madau P. \&  Rees, M. J. 2001, \apj, 551, L27
\bibitem[Maoz(1993)]{maoz93} Maoz, E. 1993, \mnras, 263, 75
\bibitem[Max et al.(2007)]{max07} Max, C., E., Canalizo, G., \& de Vries, W. H. 2007, Science, 316, 1877
\bibitem[Mayer et al.(2007)]{mayer07} Mayer, L., Kazantzidis, S., Madau, P., Colpi, M., Quinn, T., \& Wadsley, J. 2007, 
Science, 316, 1874
\bibitem[Navarro et al.(1997)]{navarro97} Navarro, J. F., Frenk, C. S., \& White, S. D. M. 1997, \apj, 490, 493
\bibitem[Pesce et al.(1992)]{pesce92} Pesce, E., Capuzzo-Dolcetta, R., \& Vietri M. 1992, \mnras, 254, 466
\bibitem[Pretorius(2005)]{pretorius05} Pretorius, F. 2005, Phys. Rev. Lett., 95, 121101
\bibitem[Richtsone et al.(1998)]{richtsone98} Richtsone, D., et al. 1998, Nature,  395, A14
\bibitem[Rodriguez et al.(2006)]{rodriguez06} Rodriguez, C., Taylor, G. B., Zavala, R. T., Peck, A. B., Pollack, L. K., \& Romani, R. W. 2006, \apj, 646, 49
\bibitem[Sesana et al.(2004)]{sesana04} Sesana, A., Haardt, F., Madau, P., \& Volonteri, M. 2004, \apj, 611, 623
\bibitem[Shields et al.(2009)]{shields09}Shields, G. A.; Bonning, E. W.; Salviander, S. 2009, \apj, 696, 1367 
\bibitem[Spergel et. al(2007)]{spergel07} Spergel, D. N., et al. 2007, ApJS, 170, 377
\bibitem[Stadel(2001)]{stadel01} Stadel, J. 2001, PhD thesis, University of Washington 
\bibitem[Stadel et al.(2008)]{stadel08} Stadel, J., Potter, D., Moore, B., Diemand, J., Kuhlen, M., Madau, P., Zemp, M., \& 
Quilis, V. 2008, MNRAS, submitted (arXiv:astro-ph/0808.2981)
\bibitem[Tanaka \& Haiman(2009)]{tanaka09} Tanaka, T., \& Haiman, Z. 2009, \apj, 696,1798
\bibitem[Tremaine et al.(2002)]{tremaine02} Tremaine, S., et al. 2002, \apj, 574, 740
\bibitem[Vicari et al.(2007)]{vicari07} Vicari, A., Capuzzo-Dolcetta, R., \& Merritt, D. 2007, ApJ, 662, 797
\bibitem[Volonteri et al.(2003)]{volonteri03} Volonteri, M., Haardt, F., \& Madau, P. 2003, ApJ, 582, 559
\bibitem[Volonteri \& Madau(2008)]{volonteri08} Volonteri, M., \& Madau, P.  2008, ApJ, 687, L57
\bibitem[Volonteri \& Rees(2006)]{volonteri06} Volonteri, M., \& Rees, M. 2006, \apj, 650,669
\bibitem[Zemp et al.(2009)]{zemp09} Zemp, M., Diemand, J., Kuhlen, M., Madau, P., Moore, B.,  Potter, D., Stadel, J., \& 
Widrow, L. 2009, MNRAS, 394,641
\bibitem[Zentner \& Bullock(2003)]{zentner03} Zentner, A., \& Bullock, J. 2003, \apj, 598, 49
\end{thebibliography}
\end{document}